\documentclass[useAMS,usenatbib]{mn2e}
\usepackage{epsfig}

\def\msun{{\rm {M}}_{\odot}}
\newcommand{\etal}{{et al.}~}
\newcommand{\eg}{{e.g.~}}
\newcommand{\ie}{{i.e.~}}

\def \ltsima{$\; \buildrel < \over \sim \;$}
\def \simlt{\lower.5ex\hbox{\ltsima}}            
\def \gtsima{$\; \buildrel > \over \sim \;$}
\def \gtsima{\mbox{$\; \buildrel > \over \sim \;$}}
\def \simgt{\lower.5ex\hbox{\gtsima}}            

\newcounter{cureqno}

\newenvironment{mathletters}{\refstepcounter{equation}%
    \setcounter{cureqno}{\value{equation}}%
    \edef\cur@eqn{\csname theequation\endcsname}%
    \def\theequation{\cur@eqn\alph{equation}}%
    \setcounter{equation}{0}}%

\title[Density Profiles]{Evolution of the Density Profiles of Dark Matter 
Haloes}
\author[Reed \etal] {Darren Reed,$^{1, 2}$\thanks{~~~~~~~~~~~~~~~~~~~~
Email: d.s.reed@durham.ac.uk.}  
Fabio Governato,$^{1,3}$
Licia Verde,$^{4,5}$
Jeffrey Gardner,$^6$
\newauthor
Thomas Quinn,$^1$
Joachim Stadel,$^7$ 
David Merritt,$^8$ 
and George Lake,$^9$\\
$^1$Astronomy Department, University of Washington, Box 351580, Seattle, 
WA 98195 USA\\
$^2$Institute for Computational Cosmology, Dept. of Physics, University of 
Durham, South Road, Durham DH1 3LE, UK\\
$^3$INAF, Osservatorio Astronomico di Brera, via Brera 28, I-20131 Milano, 
Italy\\
$^4$Dept. of Physics \& Astronomy, University of Pennsylvania, 209 South
33rd Street, Philadelphia, PA 19104-6396, USA \\
$^5$Dept. of Astrophysical Sciences, Princeton University, Peyton Hall, 
Ivy Lane, Princeton, NJ 08544 USA\\ 
$^6$Pittsburgh Supercomputing Center, 4400 Fifth Avenue, Pittsburgh, PA
15213, USA\\
$^7$Institute for Theoretical Physics, University of Zurich, 
Winterthurerstrasse 190, 8057, Switzerland\\
$^8$Department of Physics, Rochester Institute of Technology, 84 Lomb Memorial
Dr., Rochester, NY 14623-5603, USA\\
$^9$Department of Physics, PO Box 642814, Pullman, WA 99164 USA}

\pagerange{\pageref{firstpage}--\pageref{lastpage}}
\pubyear{2003}

\begin{document}

\maketitle

\label{firstpage}

\begin{abstract}

We use numerical simulations in a $\Lambda$CDM cosmology to model density
profiles in a set of sixteen dark matter haloes with resolutions of up to
seven million particles within the virial radius. 
These simulations allow us
to follow robustly the formation and evolution of the central cusp over a
large mass range of 10$^{11}$ to 10$^{14}$ $\msun$, down to approximately
0.5$\%$ of the virial radius, and from redshift 5 to the present, covering
a larger range in parameter space than previous works.  We confirm
that the cusp of the density profile is set at redshifts of two or greater
and remains remarkably stable to the present time, when considered in
non-comoving coordinates.

Motivated by the diversity and evolution of halo profile shapes, we fit our
haloes to the two parameter profile, $\rho \propto {1 \over
(c_{\gamma}r/r_{\rm vir})^{\gamma}[1+(c_{\gamma}r/r_{\rm
vir})]^{3-\gamma}}$, where the steepness of the cusp is given by the
asymptotic inner slope parameter, $\gamma$, and its radial extent is
described by the concentration parameter, $c_{\gamma}$ (with $c_{\gamma}$
defined as the virial radius divided by the concentration radius).  In our
simulations, we find $\gamma \simeq 1.4 - 0.08$Log$_{\rm 10}(M/M_{*})$ for
haloes of $0.01M_{*}$ to $1000M_{*}$, with a large scatter of $\Delta\gamma
\sim \pm 0.3$, where $M_{*}$ is the redshift dependent characteristic mass
of collapsing haloes; and $c_{\gamma} \simeq 8.(M/M_{*})^{-0.15}$, with a
large $M/M_{*}$ dependent scatter roughly equal to $\pm c_{\gamma}$.  Our
redshift zero haloes have inner slope parameters ranging approximately from
$r^{-1}$ (\ie Navarro, Frenk, \& White) to $r^{-1.5}$ (\ie Moore
\etal), with a median of roughly r$^{-1.3}$.  This two parameter
profile fit works well for all types haloes in our simulations, whether or
not they show evidence of a steep asymptotic cusp.  We also model a cluster
in power law cosmologies of $P \propto k^{n}$, with $n=$ (0, -1, -2, -2.7).  
Here we find that the concentration radius and the inner cusp slope are a
both function of $n$, with larger concentration radii and shallower cusps
for steeper power spectra.

We have completed a thorough resolution study and find that the minimum
resolved radius is well described by the mean interparticle separation over
a range of masses and redshifts. The trend of steeper and more concentrated
cusps for smaller $M/M_{*}$ haloes clearly shows that dwarf sized
$\Lambda$CDM haloes have, on average, significantly steeper density profiles
within the inner few percent of the virial radius than inferred from recent
observations.

Code to reproduce this profile can be downloaded from
http://www.icc.dur.ac.uk/$\sim$reed/profile.html.

\end{abstract}

\begin{keywords} galaxies: haloes -- galaxies: formation -- methods: 
N-body simulations -- cosmology: theory -- cosmology:dark matter
\end{keywords}

\section{Introduction}

The mass distribution of dark matter haloes provides a direct probe of the
nature of the dark matter particle, as the  inner structure of dark matter
haloes is particularly sensitive to the dark matter properties.
For example, warm dark matter should produce lower density halo
cores than cold dark matter (CDM) because of the phase density ceiling
introduced by the non-zero thermal velocity of warm particles (\eg
Tremaine \& Gunn 1979).  The variation of peak halo phase density with
halo mass is also dependent on the ``coldness'' of the dark matter
particle (\eg Lake 1989 and references therein). Spectroscopic
observations of stellar motions in galaxies, lensing properties and X-ray
temperature maps of cluster cores each can provide a measurement of the
central dark matter distribution in haloes, albeit with some uncertainty
inherent in inferring the dark matter distribution from properties of
baryons or baryon dominated regions.  CDM haloes in N-body simulations
consistently have a steep central cusp where the density rises as
$r^{-1}$ (Navarro, Frenk, \& White 1996, 1997, NFW hereafter; Huss, Jain,
\& Steinmetz 1999; Power \etal 2003), $r^{-1.5}$ (Moore \etal 1998, 1999,
M99 hereafter; Taylor \& Navarro 2001; Governato, Ghigna, \& Moore 2001; 
Fukushige \& Makino 2001, 2003),
or somewhere in between (\eg Klypin \etal 2001; Fukushige, Kawai, \&
Makino 2003; Hayashi \etal 2003; Navarro \etal 2004). These numerical
findings 
appear in  conflict with the most direct observational
results.  Rotation curves of low surface brightness (LSB) dwarfs
consistently yield density profiles with nearly constant density cores
(\eg Flores \& Primack 1994; Moore 1994; Salucci \& Burkert 2000; de Blok
\etal 2001).  Studies of more luminous galaxies imply similar problems
(\eg Salucci \& Burkert 2000; Salucci 2003; Gentile \etal 2004).
The disagreement with rotation curves may indicate an
insurmountable problem with CDM models, may be due to uncertainties in
measuring accurate stellar curves at just $\sim1\%$ of the virial radii
(van den Bosch \etal 2000; van den Bosch \& Swaters 2001; see however \eg
Simon \etal 2003), or may perhaps reflect some systematic bias common to
all high resolution N-body simulations.  Alternatively, the disagreement
may be due to a problem with common assumptions made when reconstructing
mass profiles from circular velocity data (Hayashi \etal 2003).  Strong
gravitational lensing in clusters can potentially provide a direct
measurement of the halo mass profile, and indeed central mass profiles
for several lensing clusters have been calculated (Tyson, Kochanski, \&
Dell'Antonio 1998; Shapiro \& Iliev 2000;  Sand, Treu, \& Ellis 2002;
Gavazzi \etal 2003; Sand \etal 2004), but have yielded conflicting
results. Cluster density profiles inferred from Chandra
luminosity-temperature mapping have steep cusps that are inconsistent
with the flat cores observed from LSB rotation curves (Lewis, Buote \&
Stocke 2003), though this method is sensitive to models of the
intracluster gas. In sum, many observational studies suggest a flatter
profile than predicted by CDM models, but observations have not yet
converged upon a basic shape of the density profile 
(e.g., Jimenez et al.  2003).

NFW found that CDM haloes have a ``universal'' density profile that is
independent of mass, cosmological parameters, and the initial density
fluctuation spectrum with significant scatter from halo to halo,
\begin{equation} 
\rho={\rho_s \over (r/r_s)(1+r/r_s)^2}, 
\label{eqnfw}
\end{equation} 
where $r_{\rm s}$ and $\rho_{\rm s}$ is a characteristic inner
radius (the concentration radius) and inner density,
respectively. Fukushige \& Makino (1997), based on a single halo with
$\sim10^{6}$ particles, as opposed to the $\sim10^{4}$ particles of the
NFW study, found a profile with slope between $r^{-1}$ and $r^{-2}$. M99,
using results from a series of six $\sim10^{6}$ particle haloes, also found
a profile steeper than $r^{-1}$, and proposed the following profile:
\begin{equation} 
\rho = {\rho_s \over (r/r_{\rm s})^{1.5}[1+(r/r_{\rm s})^{1.5}]}. 
\label{eqm99} 
\end{equation} 
The NFW and M99 profiles are
both specific cases of a three parameter profile family proposed by
Hernquist (1990), and further developed by Zhao (1996). The highest
resolution haloes to date are a series of eight clusters, several with
$\sim20-30\times10^{6}$ particles (Fukushige, Kuwai, \& Makino 2004),
which have central slopes steeper than NFW and shallower than M99.

The level of ``universality'' of the density profile is a matter of debate
(\eg Tatitsiomi \etal 2004).  
Jing \& Suto (2000, 2002), found central density cusps of $r^{-1.1}$,
$r^{-1.3}$, and $r^{-1.5}$ for a simulated halo with cluster, group, and
galaxy mass, respectively.  A similar range of inner slope values was found
in a recent set of high resolution haloes (Hayashi \etal 2003;  Navarro \etal
2004). The central cusp is especially sensitive to flattening caused by poor
resolution or other numerical effects (\eg Moore \etal 1998), so slightly
different numerical techniques might produce significantly different density
profiles, making it sometimes difficult to compare results of different
authors.  The appearance of near universality in many previous studies could
be due to the fact that most simulations have modelled objects in the range
of galaxies to clusters; in this mass-range the effective slope of the
linear power spectrum ($n$, where $P(k)\propto k^n$) is $n \simeq$ -2,
implying cusps of roughly NFW slope (Syer \& White 1998; Subramanian, Cen,
\& Ostriker 2000).  Ricotti (2003), using a large set of haloes with $10^{4}$
to $10^{5}$ particles, found considerably flatter cusps for high-redshift
low-mass haloes, with $r^{-0.5}$ at $z\simgt10$ for $10^{8} h^{-1}\msun$.
Flat, low mass haloes match model predictions (Syer \& White 1998;
Subramanian, Cen, \& Ostriker;  2000), wherein the central slope varies as
$r^{(9+3n)/(5+n)}$. See however, Moore \etal (2001), who find a $r^{-1.3}$
cusp in a $10^{8} h^{-1}\msun$ halo at redshift four. In CDM models, any
dependence on $n$ would be manifested as a dependence on halo mass.  In the
$\Lambda$CDM model, $n$ asymptotically approaches -3 for low mass haloes. As
$n$ nears -3, $M_*$, the characteristic mass of collapsing haloes as a
function of scale factor (see section 4.3) 
diverges, so haloes of all masses collapse nearly
simultaneously. If one models halo formation as the assembly of spherically
symmetric shells of material whose density is largely determined by the
scale factor of the universe, then the density profile should be shallower
when $n$ nears -3. 

The characteristic radius, $r_{\rm s}$, indicates the size of the central
density region and is usually defined in terms of the concentration
parameter, $c=r_{\rm vir}/r_{\rm s}$.  NFW, as well as a number of other
authors found that the concentration radius decreases with halo mass, even
for power law cosmologies where $n$ is constant. This agrees with other
simulations (\eg Bullock \etal 2001a; Eke, Navarro \& Steinmetz 2001), and
also with predictions based on the halo model wherein nonlinear properties
are predicted starting from Press \& Schechter (1974) theory (\eg
Huffenberger \& Seljak 2003). The concentration dependence on mass becomes
weaker when $n$ is closer to -3 (Eke, Navarro, \& Steinmentz 2001).

\begin{table*}
\centering
\caption{Summary of our halo sample at redshift zero.  For volume renormalized runs, the mass and particle number of the central halo is
listed.  N$_{\rm p,eff}$ is
the
effective particle number based on the high resolution region for renormalized runs}
\footnotesize
\begin{tabular}{@{}llllllllll@{}}
   &   M$_{\rm Halo}$ & N$_{\rm p,halo}$ &  N$_{\rm p,eff}$ & $r_{\rm soft}(h^{-1}{\rm kpc}$) & 
$\Theta$(z$>$2) &
$\Theta$(z$<$2) & z$_{\rm start}$ & L$_{\rm box}$ ($h^{-1}$Mpc) \\
\hline
CUBEHI & 0.7-2.1$\times10^{14}$ & 0.6-1.6$\times$10$^{6}$ &
432$^{3}$ & 5 & 0.7 & 0.8 & 69&
50 & 10 clusters\\
GRP1    &  4$\times10^{13}$ & 7.2$\times10^{6}$ & 1728$^{3}$  & 0.625 & 0.5 & 0.7 & 119 & 70 & Fornax mass\\
CL1     &  2.1$\times10^{14}$ & 4.6$\times10^{6}$ & 864$^{3}$   & 1.25  & 0.5 & 0.7 & 119 & 70 & Cluster\\
GAL1    &  2$\times10^{12}$ & 0.88$\times10^{6}$ & 2304$^{3}$  & 0.469 & 0.5 & 0.7 & 119 & 70 & Milky Way\\
GRP2    &  1.69$\times10^{13}$ & 0.38$\times10^{6}$ & 864$^{3}$ & 1.25 & 0.5  & 0.7 & 119 & 70 & Group\\
DWF1    &  1.88$\times10^{11}$ & 0.64$\times10^{6}$ & 4608$^{3}$  & 0.234 & 0.5 & 0.7 & 119 & 70 & 2 Dwarfs \\
        &  1.93$\times10^{11}$ & 0.66$\times10^{6}$& & & & & & & \\
$n=0$ & 1.9$\times10^{14}$ & 0.54$\times10^{6}$& 432$^{3}$ & 2.5 & 0.5 & 0.7 & 799 & 70 & $P \propto k^{0}$\\
$n=-1$ & 2$\times10^{14}$ & 0.55$\times10^{6}$ & 432$^{3}$ & 2.5 & 0.5 & 0.7 & 269 & 70 & $P \propto k^{-1}$\\
$n=-2$ & 1.6$\times10^{14}$ & 0.45$\times10^{6}$ &432$^{3}$ & 2.5 & 0.5 & 0.7 & 99 & 70 & $P \propto k^{-2}$\\
$n=-2.7$ & 2.9$\times10^{13}$ & 0.82$\times10^{5}$ &432$^{3}$ & 2.5 & 0.5 & 0.7 & 79 & 70 & $P \propto k^{-2.7}$\\
 \end{tabular}
\end{table*}

In spherically symmetric infall models, concentration should increase with
lower mass because lower mass haloes form early, when the universe is
more dense, so their central regions are assembled with comparatively
higher densities than their outer regions.  This should give them higher
characteristic central densities, or equivalently, higher concentration
parameters (\eg Eke, Navarro, \& Steinmetz 2001, and references therein).  
Eke, Navarro, \& Steinmetz (2001) and Bullock \etal (2001a) utilise
spherical infall models to predict that halo concentration should decrease
with redshift for a given mass for similar reasons. The spherical infall
model also gives an intuitive prediction for trends in central slope at a
given radius.  If halo mass is increasing rapidly with scale factor, then
a shallow central slope would be expected because accreting matter will
all have a relativity similar physical density. Similar, but perhaps more
physically motivated, are merger models (Syer \& White 1998; Subramanian,
Cen \& Ostriker 2000), which describe the evolution of the halo density
profile as a result of hierarchical halo formation.  In the merger models,
when smaller satellites merge with the main halo, their orbits will decay
via dynamical friction, and tidal stripping will add their mass to the
parent halo. The inner slope is then set by the dependence of the
concentration (or equivalently, the characteristic density) on satellite
mass, and so can be predicted from $n$.  In general, if small haloes formed
early, as is the case if $n \sim 0$, they should be dense, which should
yield steeper central slopes because their orbits can decay to smaller
radii before being tidally disrupted (Syer \& White 1998).

First works on halo mass profiles suffered from poor resolution that made it
difficult to evaluate the central slopes in haloes with high concentrations
(\ie at galaxy scales). Also, samples with higher resolution but with
limited statistics could not study trends over a large mass range, different
cosmologies, and were unable to evaluate the intrinsic amount of cosmic
scatter.  Finally, a large dynamical range and high mass resolution at high
redshift are necessary to follow the early assembly of the central part of
dark matter haloes. This is particularly difficult for small-mass haloes that
have higher concentrations, form earlier, and then need to be evolved by
several internal dynamical times.  A careful choice of the softening, force
errors, and number of timesteps is then necessary to avoid introducing
spurious numerical trends.

In this work, we improve upon previous studies by modelling a large set of
high resolution haloes and following their evolution over a wide range 
in mass and redshift parameter space.  We present a set of
high resolution simulations covering 3 orders of magnitude in mass, from
2$\times10^{11} h^{-1}\msun$ to 2$\times10^{14} h^{-1}\msun$.  This allows
us to search for potential mass dependent trends in halo profiles.  Our
highest resolution haloes have 4 million and 7 million particles within the
virial radius, respectively. Ten of our haloes are from CUBEHI, a single
simulation of a cosmological volume at uniform resolution.  This allows us
to analyse cosmological scatter in profile shapes with a uniform method not
subject to systematic uncertainties associated with differing numerical
parameters. Furthermore our haloes are resolved with several hundreds of
thousands of particles to redshift of 2 and, in a few cases, to redshift 4
or higher.

\section{Numerical Techniques} \subsection{The simulations} We use the
parallel KD (balanced binary) 
Tree (Bentley 1975) gravity solver PKDGRAV (Stadel 2001; see also Wadsley, 
Stadel, \& Quinn 2004) for
all of our numerical simulations. Initial conditions for the simulations
are set by mapping particles on to a random realisation of the mass power
spectrum, which is extrapolated to a sufficiently high redshift,
z$_{\rm start}$, that particle overdensities are safely in the linear regime.
For higher resolution within a single halo, we use a ``renormalized
volume'' technique of nested resolution regions, which has been successful
in a number of cosmological simulations (\eg Katz \& White 1993; Ghigna
\etal 1998). First, a low resolution cosmological simulation is completed.
Next, a halo of interest is identified.
To minimise sampling bias, volume-renormalized haloes are selected by
mass with the only additional constraint that they not lie within close 
proximity ($2-3 r_{vir}$) to a halo of similar or larger mass.
Then, the initial conditions
routine is run again to add small scale scale power to a region made up of
high resolution particles that end up within approximately two virial
radii of the halo centre, while preserving the original large scale random
waves.  This process is iterated in mass resolution increments of a factor
of eight until the desired resolution is achieved.  We have verified that 
the high-resolution haloes are free from significant contamination by massive
particles.

All of the simulations model a $\Lambda$CDM cosmology with $\Omega_m=$0.3
and $\Lambda=$0.7. We normalize the density power spectrum of our initial
conditions such that $\sigma_{\rm 8}$ extrapolated to redshift of zero is
1.0, consistent with both the cluster abundance (see \eg Eke, Cole, \& Frenk
1996 and references therein) and the WMAP normalization (\eg, Bennett \etal
2003; Spergel et al. 2003). We use a Hubble constant of $h=$0.7, in units of
100 km s$^{-1}$ Mpc$^{-1}$, and assume no tilt (i.e. a primordial
spectral index of 1). To set the initial conditions, we use the Bardeen
\etal (1986) transfer function with $\Gamma=\Omega_{\rm m}(z=0) h$.  See
Reed \etal (2003)  for further details on the uniform resolution CUBEHI run.

Our high resolution simulations are listed in Table 1. For the
volume-renormalized runs, we list the effective particle number of the
highest resolution region rather than the actual particle number.  
Softening choices are chosen based on empirical studies (\eg Moore \etal
1998; Power \etal 2003). Force softenings are $r_{\rm soft} = 5 h^{-1}$kpc
for the uniform resolution CUBEHI run, and $r_{\rm soft} = 1.5 \%$ times the
mean inter-particle spacing for the volume renormalized runs.  Long range
forces are calculated by hexadecapole expansion of the potentials of distant
tree nodes (or ``cells'') that subtend an angle less than the cell opening
angle, $\Theta$, chosen to be consistent with tests by Stadel (2001).  
$\Theta$ is set to be smaller at high redshift, when force errors due to
long range gravitational forces can have a larger contribution to total
forces since the density field is more uniform.  Timesteps for the CUBEHI run
are constrained to $\Delta t < 0.2\sqrt{r_{\rm soft}/a}$, where $a$ is the
magnitude of the acceleration of a given particle, for the CUBEHI run, and
$\Delta t < 0.175\sqrt{r_{\rm soft}/a}$ for all other runs.  These timesteps
are consistent with convergence tests for variable timestep runs 
by Power \etal (2003), where $\Delta t \simlt 0.2\sqrt{r_{\rm soft}/a}$
is found to be sufficient for
the central regions of haloes.  The number of
periodic replicas, $n_{\rm r}$ is 1 for all simulations; ${\rm n_r}$ 
determines the number of copies of the box
for gravity calculations, and hence the accuracy of the periodic force.  
Starting redshift, $r_{\rm soft}$, $\Theta$, and ${\rm n_r}$ are all tested 
for the CUBEHI run in Reed \etal (2003).

\begin{figure}
\begin{center}
\epsfig{file=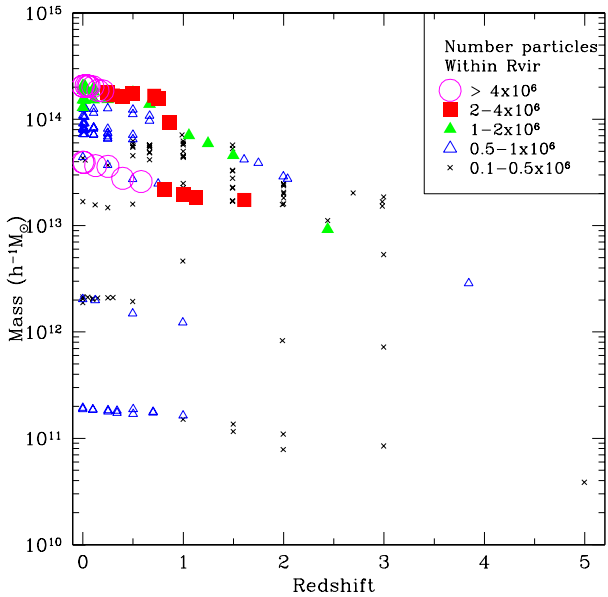, width=\hsize}
\caption{Our set of haloes for density profile
analyses, including lower resolution tests. }
\label{prosmassvsz}
\end{center}
\end{figure}

For our 
volume-renormalized 
galaxy, group, and cluster runs, we have
analysed lower resolution runs which we use for a numerical resolution
study. In the interest of limiting our study to only the highest
resolution haloes, we only calculate density profiles for the ten most
massive haloes in our CUBEHI run, each of which has $\sim$10$^{6}$
particles. For similar reasons we only follow the evolution of haloes to
redshifts where their particle number exceeds 10$^{5}$.  In Fig.  
\ref{prosmassvsz}, we plot the properties of the full sample of haloes that
we analyse, including our low resolution versions of the renormalized runs
GAL1, GRP1, and CL1 when they exceed 10$^{5}$ particles. The $n=0, -1, -2,
\& -2.7$ runs are used to study the effects of power index spectral slope
on halo structure.  The initial power spectrum for these runs is given by
$P \propto k^{n}$ normalized to the same $\sigma_{\rm 8}$ as all other runs.  
These power spectra are based on the same random waves as run CL1.

\subsection{The analysis} Our virialized haloes are selected with the (SO)
algorithm (Lacey \& Cole 1994) utilising the spherical tophat model of Eke,
Cole, \& Frenk (1996) in which the $\Lambda$CDM virial overdensity, {\it
$\Delta_{\rm vir}$}, in units of critical density is approximately 100. To
follow the evolution of an individual halo, we ``mark'' a few hundred
particles at the density peak of the halo at $z=0$ and trace those particles
back to higher redshifts, when they are in the core of the largest
progenitor haloes.  When there are multiple progenitors, we use the
progenitor with the deepest potential.

To calculate density profiles without excessive particle noise, we developed
a novel kernel-based 
algorithm\footnote{Code available at 
http://www.rit.edu/$\sim$drmsps/inverse.html}
 (Merritt \& Tremblay 1994); see Appendix for a full description. Both
the width and shape of the kernel are varied with radius; the variation in
shape is significant near the origin, where a symmetric kernel would
``overflow'' the $r=0$ boundary. 
The window width must be carefully chosen to reduce Poisson noise
(``variance'') without oversmoothing ("biasing") the profile.  In general,
it may be shown (e.g. Scott 1992, p. 130) that when the window width is chosen
to minimise the mean square error of the estimate, most of the error will
come from the variance.  Window widths large enough to eliminate the 
``wiggles''
will generally bias the slope. In addition, the window width should
vary with local particle density (Abramson 1982), roughly as $\rho^{-1/2}$.
We used a kernel window width that varied as
r$^{0.5}$ set at $h_{0.1}=.005r_{\rm vir}$ at
0.1$r_{\rm vir}$ as it yields profiles and profile slopes in good agreement
with binned profiles created with TIPSY\footnote{TIPSY is available 
from the University of
Washington N-body group: http://hpcc.astro.washington.edu.}), 
and it preserves the central cusp and major substructure.

\begin{figure*}
\epsfig{file=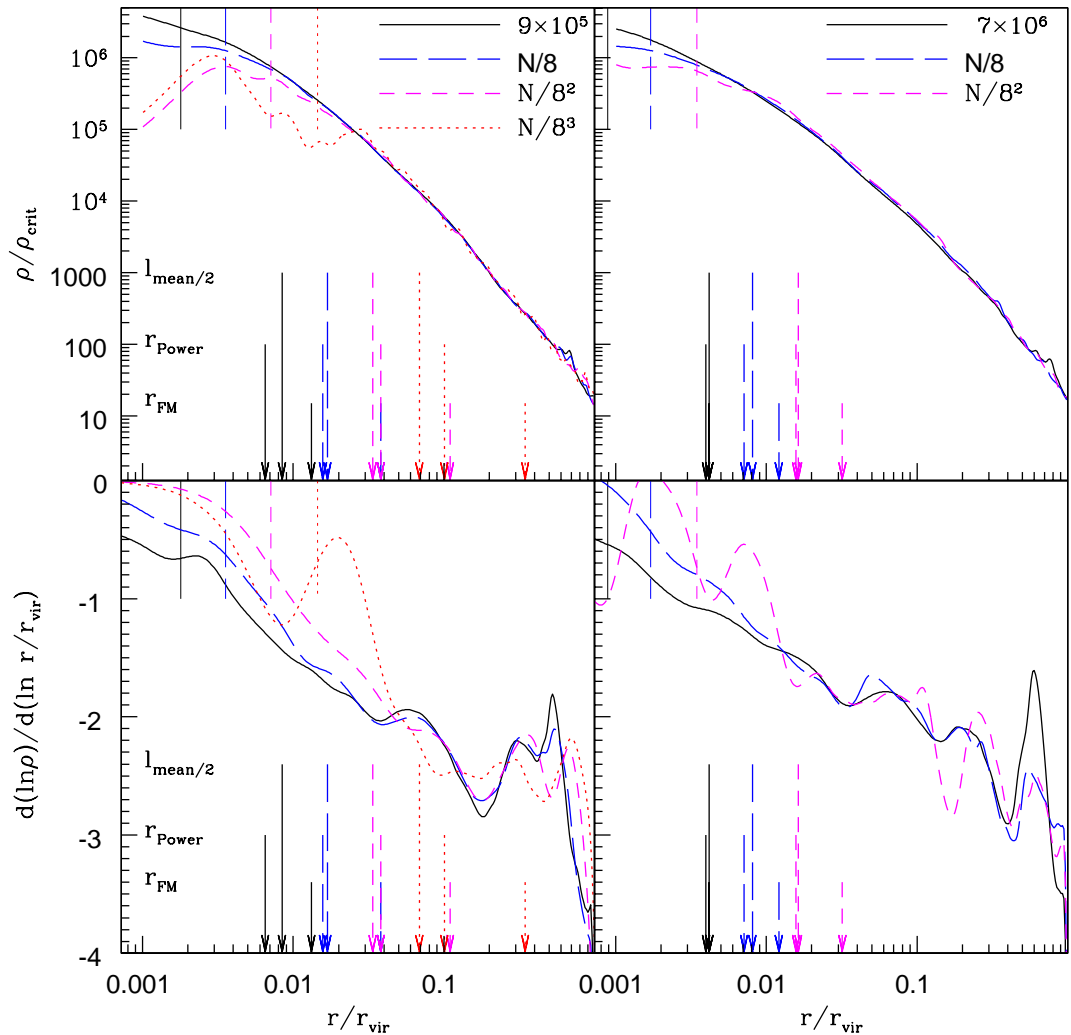, width=15cm}
\caption{Top panels: Density profiles of our galaxy halo (GAL1, left) and
our highest resolution halo (GRP1, right)
at multiple resolutions.  Long arrows show resolution criteria from
Moore \etal (1998); medium arrows show Power \etal (2003) criteria; 
and short arrows show Fukushige \& Makino (2001) criteria.  Softening is 
indicated by vertical lines at top of plot windows.  Profiles are 
constructed via a kernel estimator (Appendix).  Bottom panels: 
Derivative of the density profile of above haloes at the same 
resolutions.  Density slopes are calculated using a rolling average of 
$\Delta {\rm r/r_{\rm vir}}\simeq\pm30\%$.}
\label{tileres}
\end{figure*}

\section{Resolution Criteria}

In this section, we utilise lower resolution versions of our
renormalized-volume simulations to examine several resolution criteria.
Previous authors have proposed empirical and theoretical criteria to define
the minimum radius at which the density profile can be considered to be
``resolved'' (see Diemand \etal 2004 and references therein, whose
discussion we summarise here). The main numerical issue is the discreteness
caused by the fact that N-body particles are extremely massive compared to
dark matter candidates in the ``real'' universe.  This discreteness means
that particles will undergo two body interactions, which change their
velocity by a significant amount. Two body relaxation effects vanish as
$N_{\rm p}$ approaches infinity.  The two body relation time, defined as the
time it takes for particle energy to change by order unity, is shortest near
halo centres where density is high.  In CDM haloes, two body interactions
tend to add energy to the low velocity cusp particles, so two body
relaxation has the effect of flattening the inner cores of simulated haloes.  
After several Hubble times, haloes become nearly isothermal, and the energy
transport reverses direction. Power \etal (2003) and Fukushige \& Makino
(2001) have considered the relaxation rate at $z=0$ and found that haloes are
resolved down to radii where the relaxation time is equal to the Hubble time
and three Hubble times, respectively. Moore \etal (1998) and Ghigna \etal
(2000) offer an empirical fit finding that the minimum resolved radius is
$r_{min} \simeq l_{mean}/2$ where $l_{mean}=(4\pi/3)^{1/3}N_{\rm p}^{-1/3}$, 
based on simulations of identical haloes at
different resolutions; see Splinter \etal (1998) and references therein for
discussion relating resolution to mean particle spacing.  Because most two
body relaxation occurs early, when particles are in small haloes, particles
in higher-mass haloes can suffer significantly more 
relaxation (Diemand \etal 2004).  This is due to the later formation time 
of massive haloes, which means that their particles 
have spent more time in small $N_{\rm p}$ progenitors where two-body relaxation
is larger (Diemand \etal 2004).
We therefore test our resolution criteria over a range of masses.

We use identical haloes of varying resolution to empirically identify radii
we consider to be well resolved and compare our results with other
resolution studies.  Fig. \ref{tileres} shows the density profiles of our
galaxy and group for which we have multiple resolutions and indicates the
resolution criteria from the above studies.  We have made similar resolution
studies of cluster CL1.  In each case, none of the three resolution 
under-estimates the radii of divergence between the 
point of divergence between lower and higher
resolution haloes over the full mass range.
However, the Power \etal and Fukushige \& Makino criteria 
are more conservative for our lower resolution haloes, as they scale more
steeply with particle number.  
Fig.  \ref{tileres} also shows the slopes of
the density profiles, $d \ln\rho /d \ln (r/r_{\rm vir})$, for the same
haloes.  To reduce noise, profile slope is calculated based on a rolling
average of the kernel-based density profile of roughly $\pm$30$\%$ in
radius\footnote{Since the kernel density estimate is a continuous function 
of radius, a
better way to compute the derivatives would have been via analytical
differentiation of $\hat\nu(r)$, using a somewhat larger window width
to compensate for the increased variance of the derivative as compared
with the function itself (e.g. Scott 1992, p. 131). We recommend that
this procedure be followed in the future, though it has no significant 
effects on our results.}  
This slope estimation is sufficient for our purposes; however, we
note that an optimal method of obtaining low-noise profile slopes is to
compute the density derivative directly from a kernel based density profile
that was made using a larger kernel window.  The profile slopes seem more
sensitive to particle resolution, and thus appear to be accurate down to
minimum resolved radii that are $\sim50\%$ larger than inferred from the
density profiles.  
At low particle numbers, both Power \etal and 
sometimes the Fukushige \& Makino 
criteria appear to be over conservative, which suggests that because of
their steeper particle number dependence, these criteria may not be 
conservative
enough for haloes with very large $N_{\rm p}$.  All of our haloes seem well
resolved down to a radius a little larger than $l_{\rm mean/2}$.  We
thus utilise ${\rm r_{min}=N_{\rm p}^{-1/3}}$, which is 25$\%$ larger than 
$l_{\rm mean/2}$.  This empirical criterion seems to best match the 
dependence of ${\rm r_{min}}$ on particle number in our simulations.  
We note that one should not expect this criteria to be valid
for haloes with vastly different central densities or 
particle numbers than modelled here, as the relation between resolved radius
should depend on the central halo density and other physical properties
in addition to particle number.
We have performed
similar resolution tests of $z=1$ outputs, where it appears that most haloes
are resolved to slightly better than ${\rm N_{p}^{-1/3}}$, probably because
particles have had less time to undergo two body interactions. At this high
redshift, the Power \etal formula still gives a conservative resolution
limit, but the Fukushige \& Makino criteria becomes less conservative than
$l_{\rm mean/2}$ for our highest particle numbers.  In sum,
the minimum resolved radius of the density profile is well described by 
${\rm N_{p}^{-1/3}}$
for a wide range of halo masses and redshifts, so we adopt this
criteria for the remainder of the paper.

\section{Results}
\subsection{The full halo sample}

\begin{figure}
\begin{center}
\epsfig{file=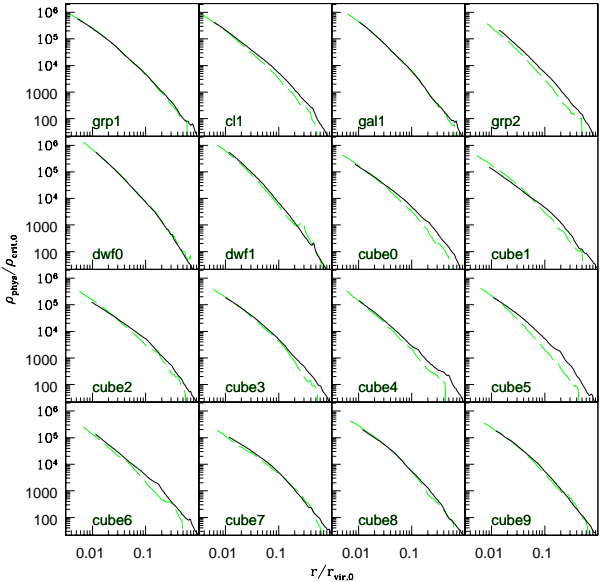, width=\hsize}
\caption{Density profiles of our 16 haloes sample for at $z=0$ (solid) and
$z=1$ (dashed).  Here we rescale the $z=1$ profiles in physical
(non-comoving) coordinates normalized to the critical density and
virial radii at $z=0$ such that a value of $r/r_{\rm vir,0}$ corresponds to
the same non-comoving distance from the halo centre at all redshifts.
Profiles are plotted to the minimum resolution
criteria of $N_{\rm p}^{-1/3}$.}
\label{protilez0z1}
\end{center}
\end{figure}

\begin{figure}
\begin{center}
\epsfig{file=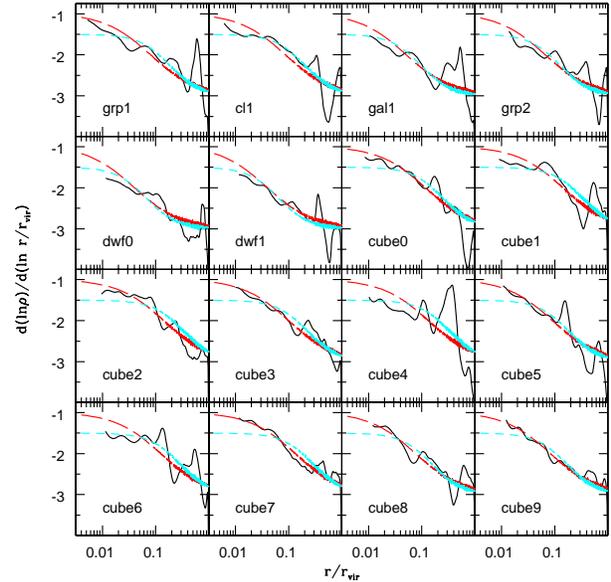, width=\hsize}
\caption{Density profile slopes of our 16 haloes sample for at $z=0$.
The best-fit concentration NFW (red, long dashed) and M99 (blue, short
dashed)
profile slopes are plotted.
Profiles are plotted to the minimum resolution
criteria of $N_{\rm p}^{-1/3}$.  Slope is calculated based on a rolling
fit of $\pm30\%$ radial width.
}
\label{slonfwmoorez0}
\end{center}
\end{figure}

In Fig. \ref{protilez0z1}, we present a plot of the density profile of all
of our haloes at redshifts of zero and one. Large substructure is apparent in
the profiles which have been constructed with the kernel algorithm described
earlier. Each halo has a unique density profile, even in the inner regions
where there is little substructure.  The cusp size and slope vary from halo
to halo.  The redshift one profiles are normalized in terms of the redshift
zero virial radius and critical density in physical (non-comoving) units,
and are plotted out to the $z=1$ virial radius.  There is little evolution
of the profile between redshift zero and one. In Fig. \ref{slonfwmoorez0},
we plot the $z=0$ slopes of the same density profiles, $d \ln\rho/d \ln
(r/r_{\rm vir})$, which is calculated based on a rolling fit of
approximately $\pm30\%$ in radius.  Substructure is prominent in the outer
regions of each halo.  Only a few of the halo density profiles appear to
converge to an asymptotic inner slope; the rest continually flatten all the
way down to the innermost resolved radii, though at a rate generally
consistent with an asymptotic slope parameter of r$^{-1}$ or steeper.  The
NFW and M99 curves are plotted for the best fit concentration parameters.
The innermost slope at the inner resolution limit ranges between r$^{-1.1}$
and r$^{-1.7}$, which implies that the halo to halo cosmic scatter in slope
is approximately bounded by the NFW and M99 profiles.

\begin{figure}
\begin{center}
\epsfig{file=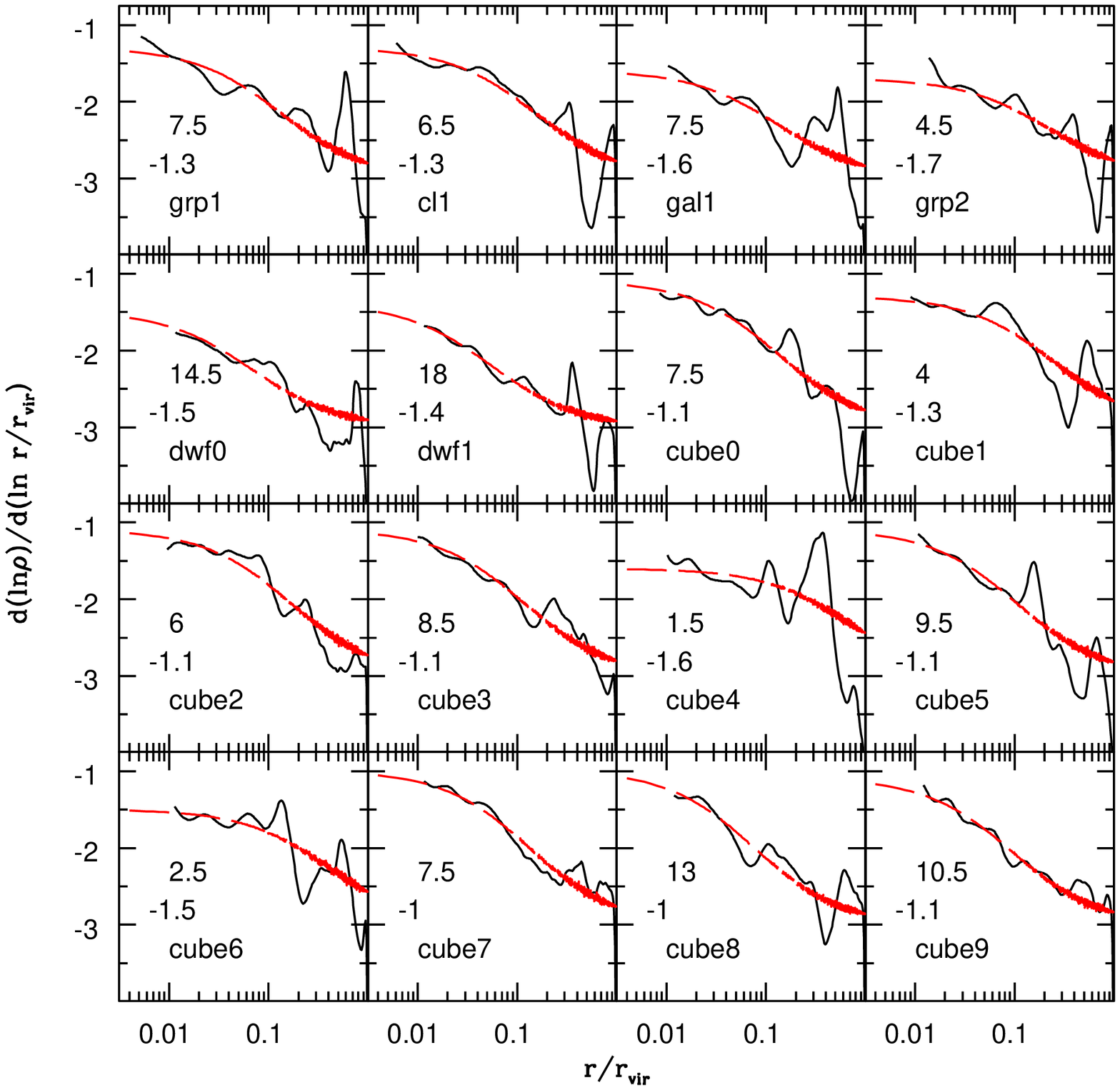, width=\hsize}
\caption{Density profile slopes of our 16 haloes sample at $z=0$.
The long dashed curve (red) shows the results of a two parameter density
profile fit in which both the inner slope
parameter, $\gamma$, and the concentration parameter, $c_{\gamma}$, are 
allowed to vary. The top number in each window is the best-fit 
$c_{\gamma}$, and the bottom is the value of the best-fit (-)$\gamma$.}
\label{slonfws}
\end{center}
\end{figure}

In Fig. \ref{slonfws}, the $z=0$ halo density slopes are plotted versus a
two parameter fit where the inner asymptotic slope parameter is allowed to
vary in addition to the concentration parameter.  The density is thus
given by:
\begin{equation}
\rho = {\rho_s \over 
(c_{\gamma}r/r_{\rm vir})^{\gamma}[1+(c_{\gamma}r/r_{\rm vir})]^{3-\gamma}},
\label{eq2param}
\end{equation}
where $\gamma$ is the asymptotic inner slope parameter and $c_{\gamma}$ is
the concentration parameter obtained when $\gamma$ is a free parameter.  
This two parameter fit produces a visually better fit to the density profile
in most cases.  The range in (-)$\gamma$ for the $z=0$ haloes is -1 to -1.7.
Note that there is a partial degeneracy between $\gamma$ and the
concentration radius. A pseudo (because the data points are correlated)
$\chi^{2}$ per degree of freedom for each of these fits shows substantial
improvement over NFW or M99 fits; see Table 2.  The profile fits are based
on a least squares method, where a Poisson uncertainty is estimated for each
point based on the effective number of particles for the density given by
the kernel-based profiles, with logarithmically spaced bins.

\begin{table*}
\centering
\begin{minipage}{140mm}
\caption{Our best-fit concentration parameters for the NFW profile (column 
3) and the M99 profile (column 5) followed
by their respective pseudo-$\chi^{2}$ per degree of freedom goodness of 
fits at redshift 0. Columns 7 and 8 are
for a two parameter fit where $\gamma$ is the central slope parameter, 
followed by its pseudo-$\chi^{2}$ in column 9.  Column 10 is the characteristic radius
for this two parameter fit}
\begin{tabular}{@{}llllllllll@{}}
halo &  r$_{\rm vir}(h^{-1}{\rm kpc}$) & c$_{\rm NFW}$ & $\chi_{\rm d.o.f.,NFW}^{2}$ & c$_{\rm M99}$ & 
$\chi_{\rm d.o.f.,M99}^{2}$ & c$_{\gamma}$ &
(-)$\gamma$ & $\chi_{\rm d.o.f.}^{2}$ &  r$_{\rm s, \gamma}(h^{-1}{\rm kpc})$\\
grp1 & 705 & 12.5 & 4.7 & 5.5 & 1.8 & 7.5 & -1.3 & 0.47 & 94 \\
cl1 & 1220 & 11.5 & 2.9 & 5 & 0.37 & 6.5 & -1.3 & 0.15 & 188 \\
gal1 & 262 & 18.5 & 1.2 & 10 & 0.98 & 7.5 & -1.6 & 0.13 & 35 \\
grp2 & 530 & 14.5 & 0.87 & 7.5 & 0.6 & 4.5 & -1.7  & 0.03 & 118 \\
dwf0 & 120 & 26.5 & 0.37 & 14 & 0.39 & 14.5 & -1.5 & 0.1 & 8.3 \\
dwf1 & 119 & 28 & 0.14 & 15 & 0.23 & 18 & -1.4 & 0.04 & 6.6 \\
cube0 & 1220 & 9 & 0.71 & 4 & 1.7 & 7.5 & -1.1 & 0.08 & 163 \\
cube1 & 1190 & 7 & 2.4 & 3 & 0.88 & 4 & -1.3 & 0.16 & 298 \\
cube2 & 1110 & 7 & 0.34 & 3 & 2.8 & 6 & -1.1 & 0.11 & 185 \\
cube3 & 1050 & 10 & 0.33 & 4.5 & 0.84 & 8.5 & -1.1 & 0.08 & 124 \\
cube4 & 1040 & 7 & 10.8 & 3 & 2.5 & 1.5 & -1.6 & 0.28 & 693 \\
cube5 & 985 & 11 & 0.27 & 5.5 & 0.63 & 9.5 & -1.1 & 0.1 & 104 \\
cube6 & 933 & 7.5 & 3.7 & 3.5 & 0.37 & 2.5 & -1.5 & 0.19 & 373 \\
cube7 & 899 & 7.5 & 0.097 & 3.5 & 2.2 & 7.5 & -1 & 0.04 & 120 \\
cube8 & 892 & 13 & 0.15 & 6.5 & 0.46 & 13 & -1 & 0.08 & 69 \\
cube9 & 863 & 12 & 0.078 & 6 & 0.27 & 10.5 & -1 & 0.02 & 82 \\
$n=0$ & 1200 & 25 & 0.47 & 13.5 & 0.54 & 14 & -1.5 & 0.14 & 85 \\
$n=-1$ & 1190 & 16 & 0.16 & 8.5 & 0.1 & 12.5 & -1.2 & 0.04 & 95 \\
$n=-2$ & 1120 & 7.5 & 0.8 & 3.5 & 0.22 & 4.5 & -1.3 & 0.06 & 249 \\
$n=-2.7$ & 635 & 6 & 0.26 & 3 & 0.51 & 5.5 & -1.1 & 0.1 & 163 \\
\end{tabular}
\end{minipage}
\end{table*}


\subsection{Redshift Evolution}

\begin{figure*}
\begin{center}
\epsfig{file=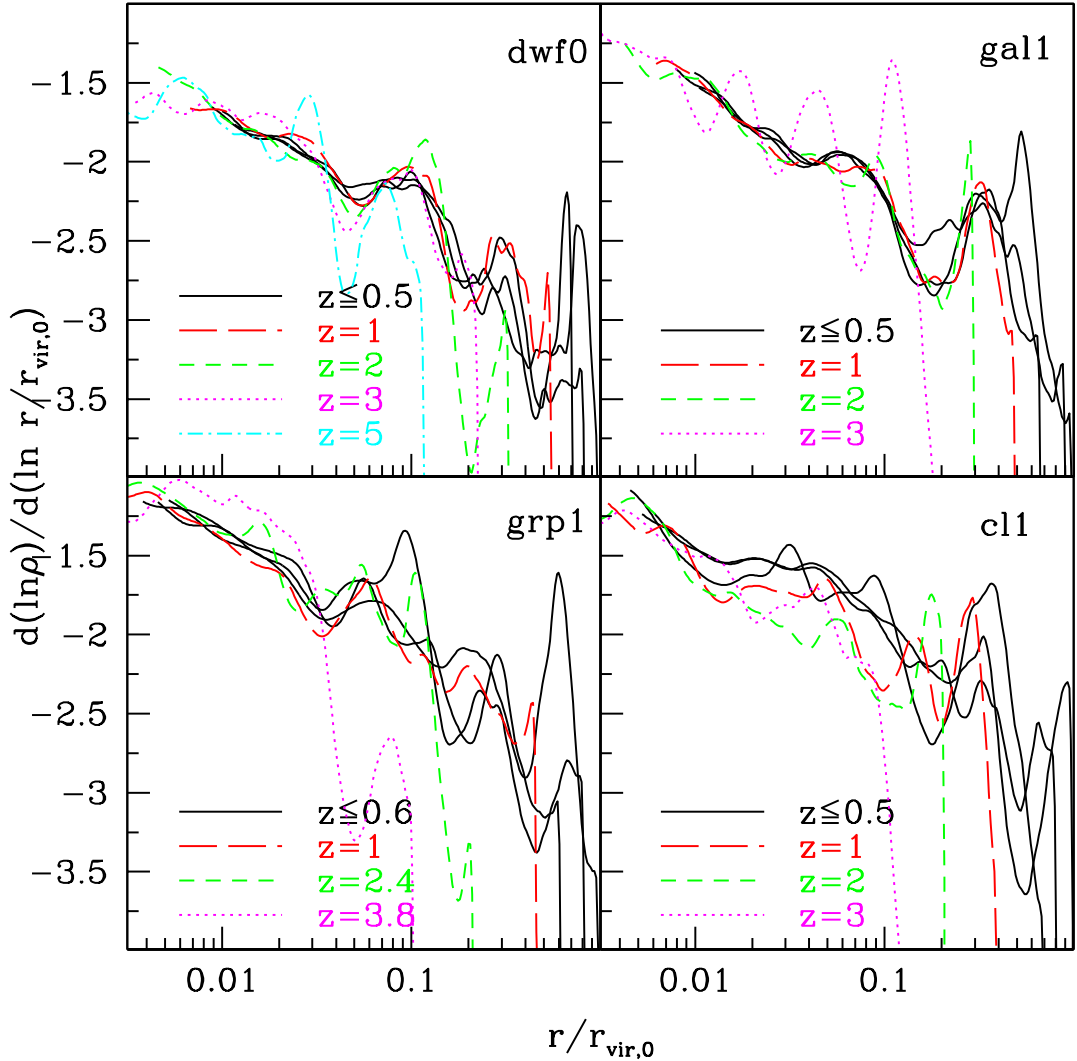, width=15cm}
\caption{Evolution of the slope of the density profiles of four haloes
plotted against $r_{\rm vir}$ at $z=0$ in non-comoving (physical) coordinates
as in Fig. \ref{protilez0z1}.
Mass beyond the virial radius of each epoch is ignored.
Note that if plotted simply in terms of
$r/r_{\rm vir}$ the haloes would appear to flatten with increased redshift.
}
\label{slotilezphys}
\end{center}
\end{figure*}

In order to gain an understanding of the physical effects that set the inner
slope of halo density profiles, we plot the evolution of the profile slope
in terms of the $z=0$ virial radius in non-comoving coordinates for our four
best haloes. This allows us to ignore the effects that expansion of the
universe or evolving $r_{\rm vir}$ have on the power law slope of the
profile.  We see in Fig. \ref{slotilezphys} that the density profile inner
slope for each halo evolves very little in non-comoving coordinates, plotted
as $d \ln\rho/d \ln (r/r_{\rm vir,0})$, a result also found by Fukushige \&
Makino (2001) and Fukushige \etal (2004).
In our cluster CL1, the slope
slowly steepens with time at $z\simgt$2, but in our other less massive
haloes, the inner density profile slope shows no significant change. The
large fluctuation in the profile slope at $z=3$ for the galaxy halo and at
$z=5$ for the dwarf halo are most likely due to the presence of subhaloes
that are disrupted earlier. Whatever physical mechanism is responsible for
setting the density profile, it must have largely occurred at very high
redshift.

The lack of evolution in the physical densities means that the profile of
an individual halo is substantially shallower at high redshift in terms of
$r/r_{\rm vir}$.  Such apparent steepening of the slope with time is merely a
scaling issue due to the growth of the virial radius as mass is added to
the outer regions of the halo.  Here we note that the profile slope is
almost never shallower than the NFW asymptotic value of $r^{-1}$. Only in
the very inner region of our largest cluster CL1 at $z=3$, where the slope
trends toward $r^{-0.75}$ at the innermost resolved radius (not plotted),
is there a slight hint that NFW slope may be too steep at high redshifts,
but this could be simply a result of a subhalo just beyond that radius
creating a local density minima, or it could be due to artificial
numerical resolution effects at high redshift where the halo resolution is
lower.


\subsection{Trends in Profile Concentration and Inner Slope}

\begin{figure}
\begin{center}
\epsfig{file=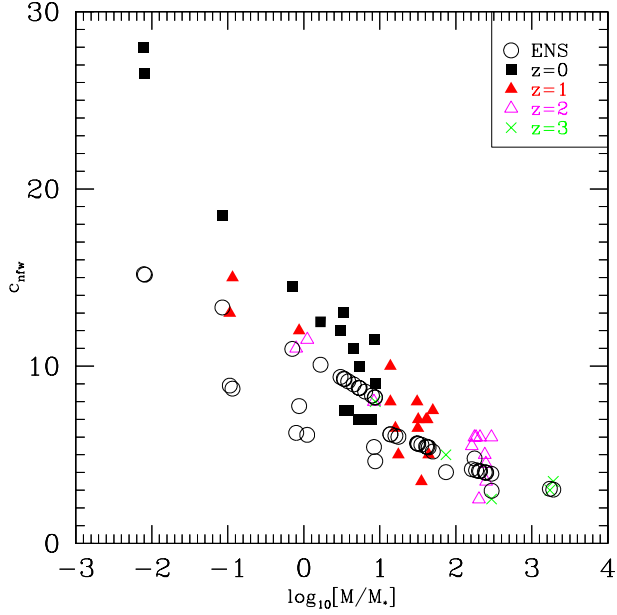, width=\hsize}
\caption{Best fit NFW concentration parameter $r_{\rm vir}/r_{\rm s}$
for our set of haloes as a function of $M/M_{*}$.  Empirical predictions by 
Eke, Navarro, \& Steinmetz (2001) are given by open circles.}
\label{conccomb}
\end{center}
\end{figure}

\begin{figure}
\begin{center}
\epsfig{file=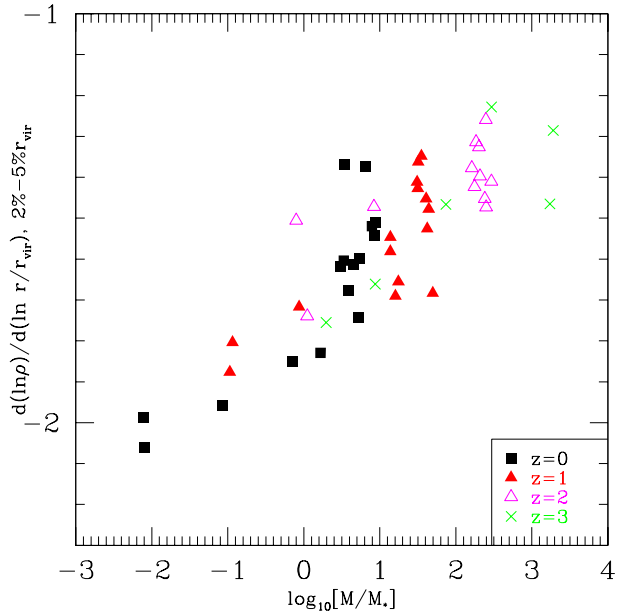, width=\hsize}
\caption{Density profile slopes of our 16 haloes sample averaged
over the range of 2$\%$ to 5$\%$ r$_{\rm vir}$.  Haloes not resolved
to 2$\%$ r$_{\rm vir}$ are not plotted.}
\label{slopemstar}
\end{center}
\end{figure}

\begin{figure}
\begin{center}
\epsfig{file=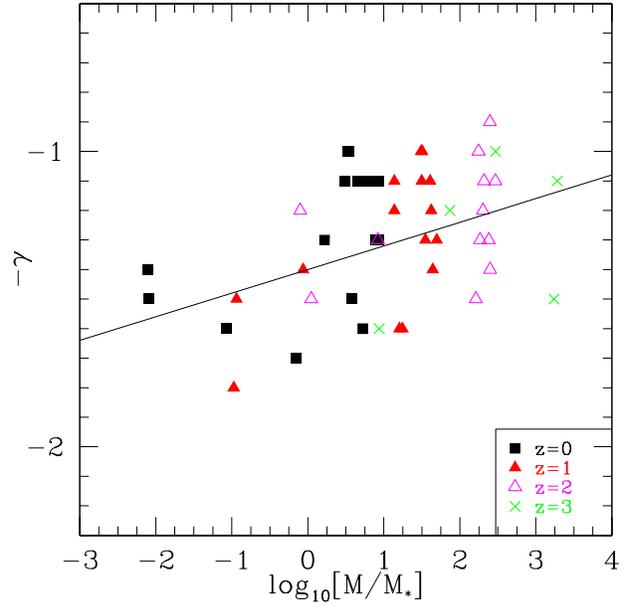, width=\hsize}
\caption{The asymptotic inner slope parameter, $\gamma$,
from a two parameter fit of our 16 halo sample.
$\gamma$ and the concentration are both allowed to vary.  Solid line fit 
is given by Eq. \ref{eqgamma}.
}
\label{gammamstar}
\end{center}
\end{figure}

\begin{figure}
\begin{center}
\epsfig{file=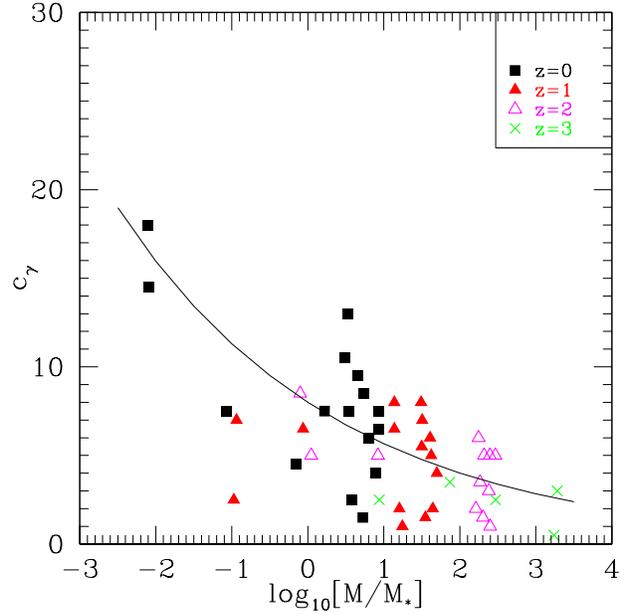, width=\hsize}
\caption{The concentration parameter from a two parameter
fit of our 16 halo sample, c$_{\gamma}$.  The asymptotic inner slope 
parameter, $\gamma$, and the concentration are both allowed to vary.
Solid line fit is given by Eq. \ref{eqconc}.
}
\label{conc2parammstar}
\end{center}
\end{figure}

The profile concentration and thus the inner slope at a given radius are
predicted by NFW and others to be a function of mass and redshift, when
considered in terms of $r/r_{\rm vir}$, with 
characteristic radius $r_s$ increasing with increasing  $M/M_{*}$.
Here, $M_{*}$ is the
characteristic mass of collapsing haloes defined by the scale at which the
rms linear density
fluctuation equals the threshold for non-linear collapse 
(\ie $\sigma(M_*(z)) = \delta_c$).
We have measured the NFW concentration parameter, $c_{\rm NFW}$, by forcing
our profiles to an NFW profile, and performing a least squares fit. In Fig.
\ref{conccomb}, we plot the concentration parameter for our set of haloes,
and we show the Eke, Navarro, \& Steinmetz (2001) prediction for $c_{\rm
NFW}$. The concentration dependence on $M/M_{*}$ for our haloes is
significantly steeper than predicted by NFW for $M<M_{*}$, and within the
scatter of our haloes for $M>M_{*}$.  Measured values of the inner slopes
also show clear trends with mass and redshift, shown in Fig.
\ref{slopemstar}, though this is due at least in part, to the degeneracy
between slope and concentration.  Here, we have measured the average value
of the power law slope $d \ln\rho/d\ln (r/r_{\rm vir})$ between 2$\%$ and
5$\%r_{\rm vir}$. The asymptotic inner slope parameter, $\gamma$, from a
two-parameter fit of $\gamma$ and concentration as given by equation 3, is
plotted in Fig. \ref{gammamstar}, and has a weak dependence on $M/M_{*}$ for
our haloes, with a large scatter. The median value of $\gamma$ in our sample
trends toward shallower inner slopes with increasing $M/M_{*}$, given by:
\begin{equation}
\gamma \simeq 1.4 - 0.08{\rm Log}_{10}(M/M_{*}),
\label{eqgamma}
\end{equation}
with a scatter of $\Delta\gamma \sim \pm 0.3$ for our haloes, and valid
for haloes of $0.01M_{*}$ to $1000M_{*}$.
Fig. \ref{conc2parammstar} shows that the concentration parameter in this
two-parameter fit, c$_{\gamma}$, shows a significant trend
toward higher values as $M/M_{*}$ decreases.  The trend is weaker and 
has significantly more scatter than the
forced NFW fit $c_{\rm NFW}$ dependence on $M/M_{*}$ (see Fig. 
\ref{conccomb}).  Adopting a power-law parameterisation as in
\eg Huffenberger \& Seljak (2003), 
the median $c_{\gamma}$ for our haloes is: 
\begin{equation}
c_{\gamma} \simeq 8.(M/M_{*})^{-0.15}, 
\label{eqconc}
\end{equation}
with a $M/M_{*}$ dependent scatter roughly equal to $\pm c_{\gamma}$.

\clearpage

\clearpage

\subsection{Cosmological Variance and Stability of Profiles} 

\begin{figure*}
\begin{center}
\epsfig{file=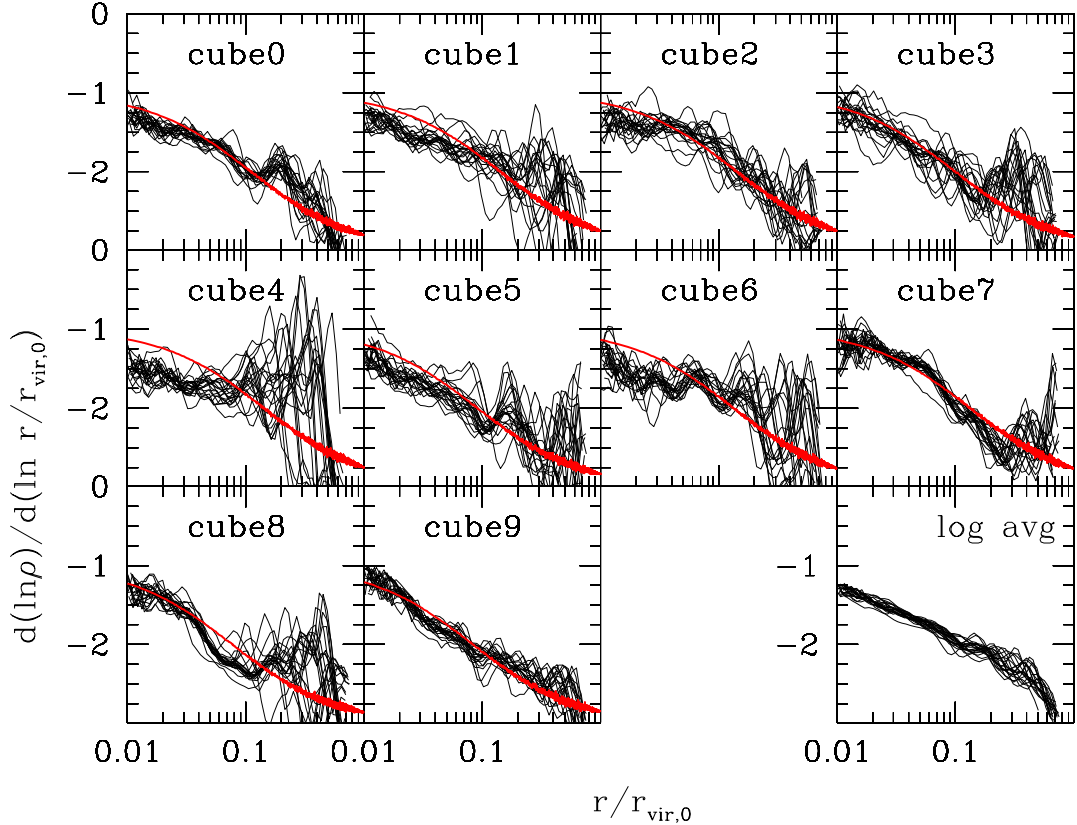, width=15cm}
\caption{Density profile slopes of our 10 clusters from the
CUBEHI simulation.  Here we plot the density profile at intervals
separated by $\Delta$z$\simeq$0.01 up to $z=$0.2.  Note that these
profiles
are binned, rather than kernel-based.  From top to bottom and left to
right, the haloes are
plotted based on descending mass.  The lower right plot is the
average of the slope value for all ten clusters.   The NFW profile
with the best fit concentration value is denoted by the solid (red) curve
for each halo
}
\label{cubeslo}
\end{center}
\end{figure*}

In our CUBEHI simulation, we examine the stability in the profiles of our
set of ten clusters over timescales separated by $\Delta z \simeq 0.01$
for $z\leq0.2$; Fig. \ref{cubeslo}.  In the outer regions, orbiting
substructure creates substantial scatter, however, the inner density
profile is relatively stable. Thus, differences between the central
density profiles of haloes of similar mass reflect different inherent
properties of the halo, rather than temporal effects of orbiting
substructure.

In the hopes of understanding what physical processes are responsible for
setting the central density profile of each halo, we examine the evolution
of the main halo and its progenitors.  We construct merger histories for
each cluster and examine a number of properties, including: evolution of the
central cusp mass concentration;  cluster accretion history; total collapsed
progenitor mass; and angular momentum profiles.  To follow the mass
accretion history of the cluster and its progenitors, we use the
friends-of-friends algorithm (FOF, Davis \etal 1985). In order to follow the
evolution of the mass concentration that makes up the cluster cusp, we use
SKID\footnote{SKID available at: http://hpcc.astro.washington.edu.} 
(Stadel 2001), which is able to identify bound mass concentrations
independently of environment.

\begin{figure*}
\begin{center}
\epsfig{file=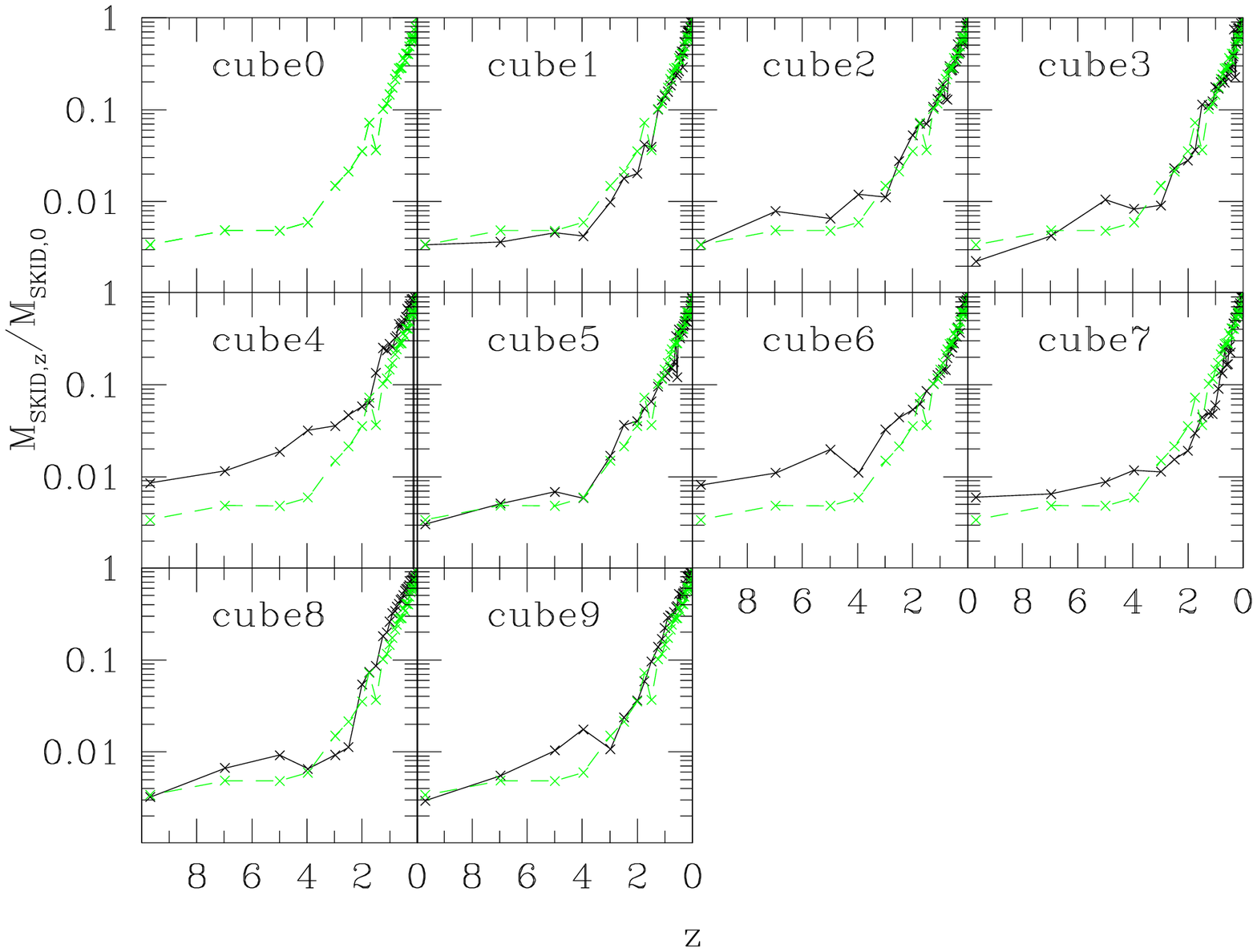, width=15cm}
\caption{Evolution of the mass of the central SKID
progenitor halo for the same ten clusters as Fig. \ref{cubeslo}.   For 
reference, the most massive halo is replotted in green (dashed curve) in 
each plot window.
}
\label{cubeskidcen}
\end{center}
\end{figure*}

\begin{figure*}
\begin{center}
\epsfig{file=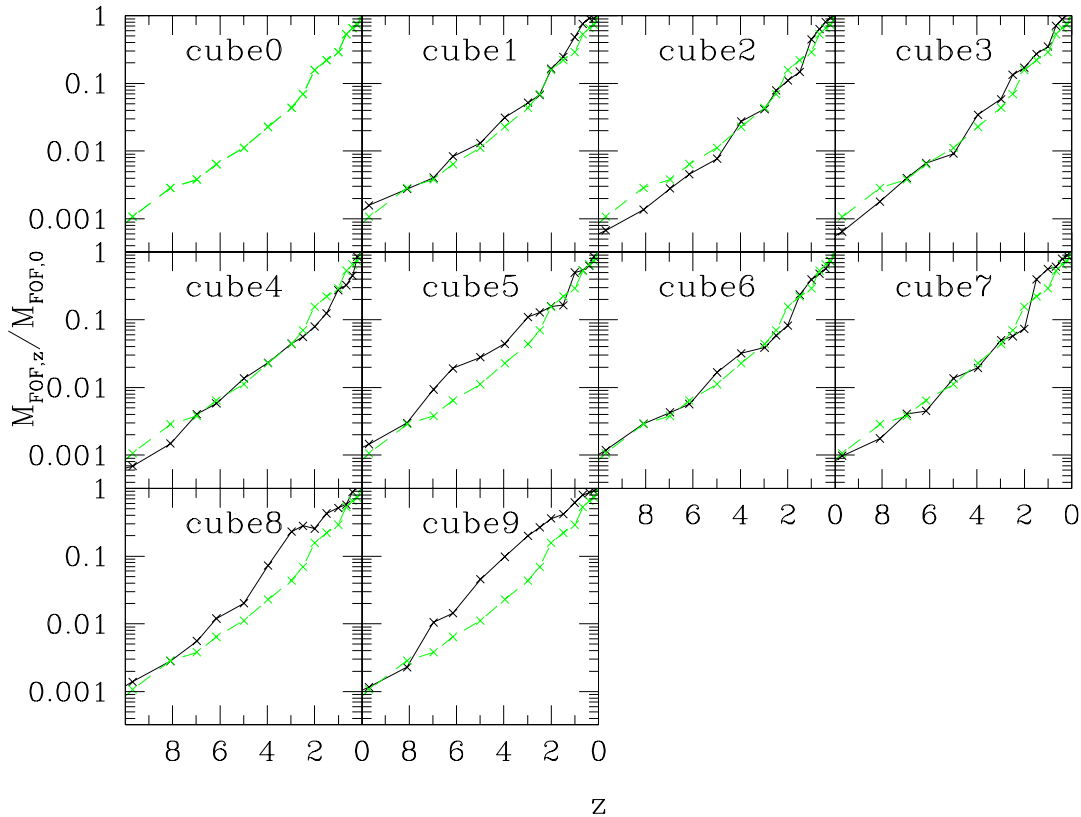, width=15cm}
\caption{Evolution of the most massive FOF progenitor halo for the same 
ten clusters as in the previous 2 figures.  For reference, the most 
massive halo is plotted in green (dashed curve) in each plot window.
}
\label{cubefofcen}
\end{center}
\end{figure*}

\begin{figure*}
\begin{center}
\epsfig{file=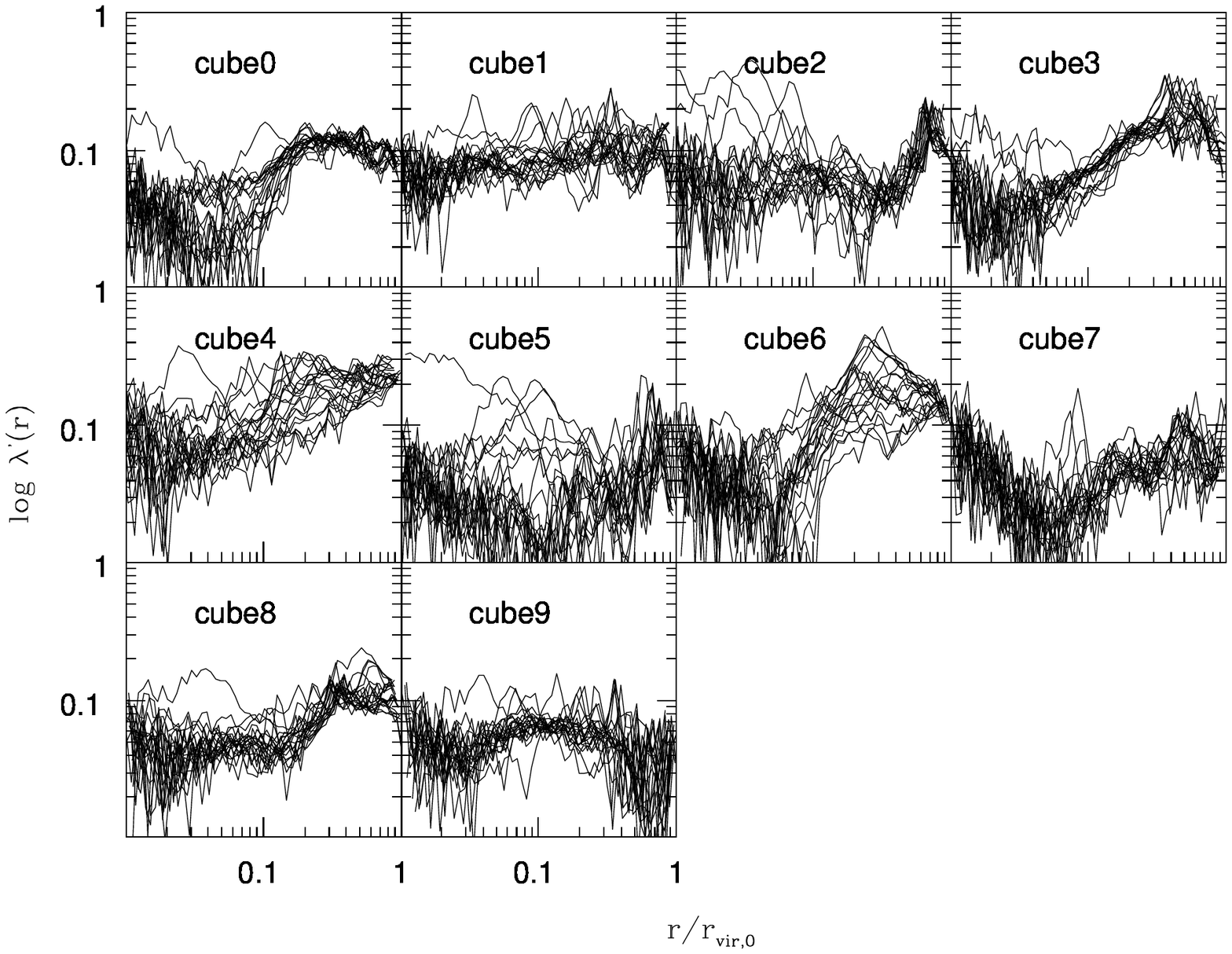, width=15cm}
\caption{Angular momentum profile of the ten CUBEHI clusters shown in the 
previous three figures.
}
\label{cubemom}
\end{center}
\end{figure*}

In Fig. \ref{cubeskidcen}, we plot the mass evolution of the central SKID
progenitor.  The two haloes with the earliest forming central SKID halo are
also those that have density profiles most closely matching the M99 profile
(cube4 and cube6; see Fig. \ref{slonfwmoorez0}, Table 2), both with slopes
that steepen slowly toward r$^{-1.5}$, beginning at radii of roughly
10$\%r_{\rm vir}$, implying large concentration radii.  If not simply a
coincidence, then this implies that the central cusp material is assembled
earlier in haloes with steeper central slope parameters.  We then consider
the effects of the mass accretion history of the cluster on the final
density profile, and find that accretion history correlates with halo
concentration, not with cusp slope.  Fig. \ref{cubefofcen} shows that the
three haloes with the highest concentration undergo a phase of rapid growth
at $z \simeq 2-8$, making their normalised mass 
temporarily $\sim 3$ times larger than
the other seven clusters, which experience nearly uniform accretion rates.  
We have also examined the evolution of the total collapsed progenitor mass,
and find a similar correlation with cluster concentration.  At redshifts of
$\sim 10$, the same three highly concentrated haloes have $\sim 3$ times
more total mass in collapsed progenitors (not plotted), where we have only
considered progenitors of mass greater than $0.01$ percent of the final
cluster mass.  The correlation of the concentration parameter with halo and
progenitor collapse time is not surprising; in fact, as have shown in
section 4.3, lower mass haloes, which are assembled earlier in a
hierarchical model, have smaller concentration radii.  The concentration
trends also qualitatively agree with correlations of formation epoch and
concentration found in numerical simulations by Wechsler \etal (2002).
However, a cautionary note is needed here.  Even though in the CUBEHI
simulation, the halo masses differ by only a factor of $\sim 3$ and thus
cover a narrow range in median concentration parameter (see eq. 5), the two
least massive and hence most poorly resolved clusters, are also the two most
concentrated.  A larger set of higher resolution haloes is needed to
conclusively rule out the possibility that haloes resolved with fewer
particles (and identical softening) lead to higher concentrations. In Fig.
\ref{cubemom}, we plot the angular momentum parameter
$\lambda^{'}=j/\sqrt{2v_cr}$, from Bullock \etal (2001b), where $j$ is the
specific angular momentum.  We see no correlation of profile concentration
or central slope with $\lambda^{'}$ in either the inner or the outer
regions.  We also have examined the angular momentum profiles at high
redshift, but find no clear correlation with halo concentration or central
slope.  This is puzzling given that merger histories, which correlate with
the density profiles, should also correlate with the angular momentum
distribution.

\subsection{Power Law Cosmologies} 
\begin{figure}
\begin{center}
\epsfig{file=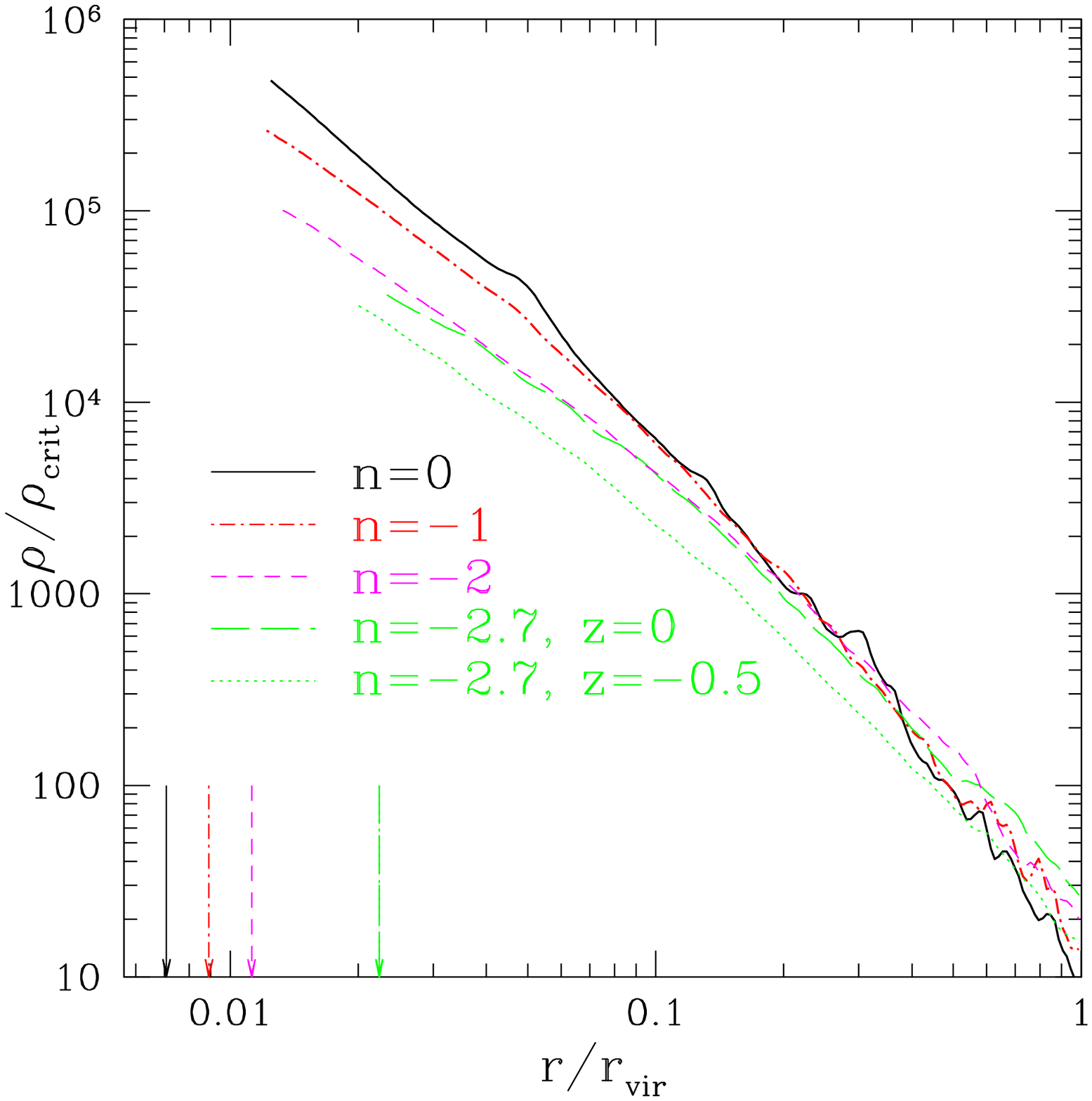, width=\hsize}
\caption{Density profiles for our cluster with initial
power spectrum given by $P\propto k^{n}$ plotted down to 
${\rm r_{min}=N_{\rm p}^{-1/3}}$ with arrows denoting the Power \etal 
resolution criteria.
}
\label{plawpro}
\end{center}
\end{figure}

\begin{figure}
\begin{center}
\epsfig{file=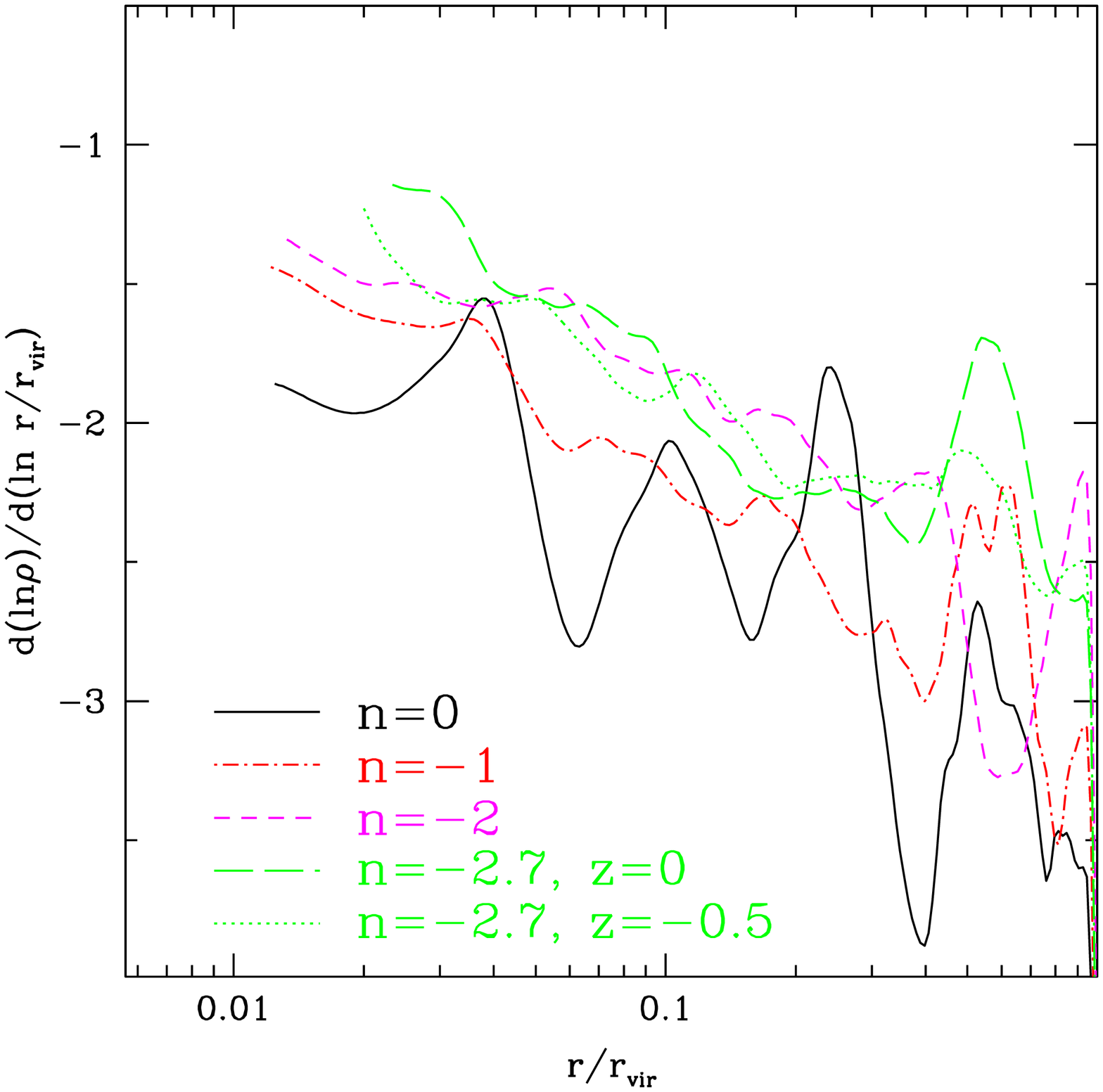, width=\hsize}
\caption{Slopes of the density profiles of Fig. \ref{plawpro}, again for
our cluster with initial
power spectrum given by $P\propto k^{n}$.
}
\label{plawslo}
\end{center}
\end{figure}

In order to understand the effects of
the power spectral slope index, $n$, we have simulated the renormalized
volume cluster CL1 in scale free cosmologies with a range of values for
$n$.  Here the initial density fluctuation power spectrum is given by $P
\propto k^{n}$, and is normalized to $\sigma_{8}=1.0$ with
$\Omega_{\rm m}=0.3$ and $\Lambda=0.7$.  Fig. \ref{plawpro} shows the density
profile for a cluster with power law initial conditions given by $n=$0,
-1, -2, and -2.7, followed by the corresponding density profile slopes in
Fig. \ref{plawslo}.  Here we also include plot the Power \etal density criteria
as we have not done convergence tests specifically for $P \propto k^{n}$ 
cosmologies.
Note that the $n=-2.7$ run has significantly less mass
because of the $\sigma_{8}=1.0$ normalization.  There is a clear trend
that steeper power spectra yield density profiles with flatter slopes and
larger concentration radii.  Visually, the $n=0$ cluster displays much
more prominent substructure, due to its proportionally stronger small
scale power, with many nearly spherical haloes and few filaments.  A two
parameter profile fit yields $\gamma=1.5$ for $n=0$ and $\gamma=1.1$ for
$n=-2.7$ (see Table 2). A number of authors (Hoffman \& Shaham 1985;  
Crone, Evrard, \& Richstone 1994; Cole \& Lacey 1996; Syer \& White 1998;  
Subramanian \etal 2000; Huffenberger \& Seljak 2003; NFW 1997; Eke,
Navarro \& Steinmetz 2001; Ricotti 2002) have suggested a power law
dependence of the density profile that qualitatively agrees with our
power law simulations.  However, Syer and White (1998) predict that an
$n=-2.7$ power law cosmology should have an inner slope of $r^{-0.4}$,
which is much shallower than seen in our haloes, though not necessarily
inconsistent if the slope flattens only at very small radii. The same
model predicts an inner slope of $r^{-1.8}$ for the $n=0$ cosmology,
which is very close to the inner slope at the minimum resolved radius of
our $n=0$ halo.

\section{Discussion and Conclusions}

(1) {\it Density profile trends and the CDM ``cusp problem''.} We follow the
evolution of sixteen haloes over a large range in parameter space,
with masses corresponding to dwarfs through
clusters back to $z=3$ for our highest resolution cluster and $z=5$ for two
dwarfs.  By using identical haloes at varying resolution, we have shown that
our haloes are likely to be free from biases related to numerical resolution.  
None of our simulated haloes has a density profile slope significantly
shallower than $r^{-1}$ down to the minimum resolved radii of 0.5$\%$ to
2$\%$. These steep cusps are similar to those found in previous CDM
simulations, and appear to be in conflict with reported rotation curves from
dwarfs and LSBs.  Our resolution experiments confirm that steep cusps are
formed regardless of numerical resolution, down to the minimum resolved
radius of each halo.  Previous high resolution simulations discussed earlier
have mostly modelled haloes with masses of $10^{12} h^{-1}\msun$ or higher,
leaving the possibility that a mass dependence on the density profile might
be able to solve the cusp problem. However, our results show that the
observed conflict with $\Lambda$CDM density profiles gets worse when
considering simulated dwarfs.  The enclosed mass at a given radius near the
halo core is higher for simulated haloes of lower mass, since the
concentration radius is smaller and the measured slope is steeper.

(2) {\it Convergence of profile slopes.} We confirm that the physical slope of
haloes remains stable over time.  The inner regions of haloes appear to be
composed of mass assembled at very high redshift, implying that 
present-epoch density-profiles are determined by the high redshift
 merger history. We do see some apparent evolution in the physical slope at
very high redshifts for our highest resolution cluster and group.  This
suggests that for group and clusters-size haloes, events occurring at $z
\simeq 2-4$ are partially responsible for determining the final shape of the
density profile, while galaxies and smaller mass haloes have their inner
density profile shape almost entirely determined at higher redshifts.  The
progenitor region of the host halo, because of its large scale density
enhancement, is the likely site of early-forming small haloes, many of which
will merge to form the cusp. After the universe has expanded by a few
factors since the cusp material is assembled, the characteristic densities
of haloes merging into the main halo should be lower than typical densities
within the main cusp, because of their generally later assembly epoch.
Consequently, merging subhaloes will likely be disrupted before dynamical
friction can bring them near the centre of the main halo, where they can
affect the central density profile. In this scenario, halo density profiles
could converge to roughly flat profiles (in terms of $r/r_{\rm vir}$) at
extremely high redshifts.  CDM haloes at low redshift might then be expected
to have a very small flat inner core, but at radii much less than the inner
$\sim1\%$ that can be currently probed by simulations or observations.

(3) {\it Lessons from the power law cosmologies.} The dependence of the
concentration parameter on $M/M_{*}$ is most likely the result of 
a mechanism analogous to what sets the dependence of concentration 
on the spectral slope in our power law cosmologies. 
 In the power law cosmologies, the ratio of local small-scale to large-scale
power depends primarily on the spectral index.  When there is lots of small
scale power, the central density profile is generally steeper at any
specific radius than when there is little small scale power.  In the case of
a shallow spectral index, subhaloes form early, which implies that they have
small characteristic radii and thus large characteristic densities, whereas
steep spectral indices have late-forming subhaloes.  Similar formation epoch
arguments can also explain the steeper profiles of small $M/M_{*}$ haloes.
For smaller $M/M_{*}$ haloes, many subhaloes should form earlier relative to
their hosts, since the subhaloes lie at lower $\sigma$ fluctuations in the
density field relative to the local fluctuation on the scale of the host
halo.  Alternatively, smaller $M/M_{*}$ haloes, which generally form earlier,
and hence when the mean density of the universe is evolving more rapidly,
will have subhaloes and cusp material assembled over a wider range of
densities than for large $M/M_{*}$ haloes. Haloes with small $M/M_{*}$ are
thus similar to haloes with a shallow spectral index, because they have
subhaloes that form earlier at higher densities relative to the host halo. In
models where the inner slope is formed from disrupted subhaloes infalling via
dynamical friction, lots of small-scale power should give the main halo a
larger concentration parameter.  In fact, when there are lots of dense
subhaloes more of them will be able to reach closer to the core, steepening
the halo nearer to the centre.  

(4) {\it Is there an asymptotic cusp?} A strength of the NFW profile, is
that varying just the concentration radius can yield reasonable fits to
haloes with very different inner slopes at scales currently resolved by
simulations.  However, there are no compelling theoretical arguments that
the halo profile converges to a slope of $r^{-1}$ at smaller radii. The
analytical models mentioned in the introduction imply that the core slope
should be a function of $M/M_{*}$ or power spectral slope.  It is possible,
and theoretically motivated, that haloes of higher $M/M_{*}$ or steeper
spectral slope than what we have simulated here would actually have a
shallower slope than $r^{-1}$, which would be inconsistent with the NFW
profile.  Most of our haloes have density profiles that get ever shallower
with decreasing radius down to our minimum resolved radius, although they
are still consistent with asymptotic cusp slopes of $r^{-1}$ or steeper.
Additionally, a few of our haloes appear to have converged to slopes at
$r^{-1.5}$ or steeper, which would be inconsistent with the NFW profile,
though because of the degeneracy between central slope and concentration
parameter, the NFW profile is not definitively ruled out.  Simulations of
haloes with very high $M/M_{*}$, very steep power spectral indices, or
orders of magnitude more particles able to probe much farther inward, should
be able to test whether or not the r$^{-1}$ NFW central slope corresponds to
a minimum possible slope set by the physical processes of CDM halo
formation.  Navarro \etal (2004) recently proposed a new profile form based
on fits to a set of high resolution haloes.  The Navarro \etal profile
differs from ours in that it has no asymptotic cusp, but instead continually
flattens with decreasing radius. It does not become shallower than r$^{-1}$
until radii smaller than currently resolvable, so it is generally consistent
with most of our haloes.  However, our profile form is more flexible in that
it is able to better match haloes that have steep asymptotic cusps with
large concentration radii, and is still a good match to those haloes that
continuously flatten down to the minimum resolved radii in our simulations.
Future simulations will likely need to probe below $\sim$0.001 r$_{\rm vir}$
in order to determine whether dark haloes have an asymptotic central cusp.  
The mass distribution at such small radii has a strong effect on the flux of
hypothesized dark matter annihilation signals that may be detectable via
$\gamma$-rays from the galactic centre (\eg Stoehr \etal 2003; Evans \etal
2004).

(5) {\it Profile scatter and the two parameter profile} When we plot our
haloes with the best fit concentration parameters, neither the NFW nor the
M99 function provide a good fit to all of the haloes.  However, because
each halo has a unique profile determined by poorly understood and complex
high redshift events, any possible single parameter profile will be
unlikely to fully describe all haloes. Our two parameter fit describes the
general trends with $M/M_{*}$, providing a significant improvement over
NFW and M99, but it still does not account for the large halo to halo
scatter at a given mass and redshift. Our evidence in the CUBEHI clusters
for a likely correlation between the cusp material assembly epoch and cusp
slope, and between halo accretion history and concentration radius,
suggests that the density profiles might be at least partly determined
from the evolutionary history of the progenitor haloes and substructures
clumps.  Larger sets of high resolution haloes should be able to quantify
the role of mergers and other stochastic processes in shaping the density
profile.

\section*{Acknowledgements} We are grateful to the anonymous
referee for helpful and constructive comments and suggestions.
We thank Lucio Mayer for assistance with one of
the runs.  DR has been supported by the NASA Graduate Student Researchers
Program.  DR was partially supported by PPARC. FG is a David E. Brooks
Research Fellow.  FG was partially supported by NSF grant AST-0098557 at the
University of Washington. TRQ was partially supported by the National
Science Foundation. Simulations were performed on the Origin 2000 at NCSA
and NASA Ames, the IBM SP4 at the Arctic Region Supercomputing Center (ARSC)
and at CINECA (Bologna, Italy), the NASA Goddard HP/Compaq SC 45, and at the
Pittsburgh Supercomputing Center. We thank Chance Reschke for dedicated
support of our computing resources, much of which were graciously donated by
Intel.

{}


\appendix
\section{Kernel Routine for Density Profile Estimation}

Here we present the algorithms which we used to derive smooth
estimates, $\hat\nu(r)$ and $\hat\Sigma(R)$, of the particle number
density and surface density profiles from the $N$-body positions.

The routines in {\sf MAPEL\ } (Merritt 1994) allow one to derive maximally
unbiased estimates of $\nu$ and $\Sigma$ using penalty functions that embody
the approximate power-law nature of these functions.  However the {\sf
MAPEL\ } routines are relatively slow, and this fact presented difficulties
when constructing estimates using the $N\sim 10^6$ particle data sets.  
Kernel based algorithms are faster but potentially more biased; however we
found that density profiles produced with {\sf MAPEL\ } and kernel methods
are in excellent agreement down to our mininum resolved radii.  Thus we have
used the kernel estimator in our analyses.

Our derivation follows that in Merritt \& Tremblay (1994).
In the absence of any
symmetries in the particle distribution, a valid estimate of the
number density $\nu$ corresponding to a set of particle positions
${\bf r}_i$ is
\begin{equation}
\hat\nu({\bf r}) = \sum_{i=1}^N {1\over h^3} K\left[{1\over h}
\left|{\bf r} - {\bf r}_i\right|\right]
\end{equation}
where $h$ is the window width and $K$ is a normalized kernel,
e.g. the Gaussian kernel (see Fig. A1):
\begin{equation}
K(y) = {1\over(2\pi)^{3/2}}e^{-y^2/2}.
\end{equation}

Now imagine that each particle is smeared uniformly around the surface
of the sphere whose radius is $r_i$ and whose origin is at (0,0,0).  If the 
density profile is
actually spherically symmetric, this smearing will leave the density
unchanged; if not, it will produce a spherically symmetric
approximation to the true profile.  The spherically-symmetrized
density estimate is
\begin{mathletters}
\begin{eqnarray}
\hat\nu(r) &=& \sum_{i=1}^N {1\over h^3} {1\over 4\pi} \int d\phi \int
d\theta\ \sin\theta\ K\left({d\over h}\right), \\
d^2 &=& \left|{\bf r} - {\bf r}_i\right|^2 \\
&=& r_i^2 + r^2 - 2rr_i\cos\theta
\end{eqnarray}
\end{mathletters}
where $\theta$ is defined (arbitrarily) from the ${\bf r}_i$-axis.
This may be written in terms of the angle-averaged kernel $\tilde{K}$:
\begin{mathletters}
\begin{eqnarray}
\hat\nu(r) &=& \sum_{i=1}^N {1\over h^3} \tilde{K}(r,r_i,h),\\
\tilde{K}(r,r_i,h) &\equiv& {1\over 4\pi} \int d\phi \int
d\theta\ \sin\theta\ K \nonumber\\
&& \times \left(h^{-1}\sqrt{r_i^2 + r^2 - 
2rr_i\cos\theta}\right)\\
&=& {1\over 2} \int_{-1}^1 d\mu\ K\left(h^{-1}\sqrt{r_i^2 + r^2 -
2rr_i\mu}\right).
\end{eqnarray}
\end{mathletters}
Substituting for the Gaussian kernel, we find
\begin{eqnarray}
\tilde{K}(r,r_i,h) = {1\over (2\pi)^{3/2}} \left({r r_i\over
h^2}\right)^{-1}  \nonumber\\
\hspace{1in} \times \hspace{.1in} e^{-(r_i^2+r^2)/2h^2} \sinh(rr_i/h^2).
\end{eqnarray}
A better form for numerical computation is
\begin{eqnarray}
\tilde{K}(r,r_i,h) = {1\over 2(2\pi)^{3/2}} \left({r r_i\over
h^2}\right)^{-1}  \nonumber\\
\hspace{1in} \times \hspace{.1in} \left[e^{-(r_i-r)^2/2h^2} - 
e^{-(r_i+r)^2/2h^2}\right].
\end{eqnarray}

We want to vary the window width with position in such a way
that the bias-to-variance ratio of the estimate is relatively
constant.  Let $h_i$ be the window width associated with the $i$th
particle.  The density estimate based on a variable window width is
\begin{equation}
\hat\nu(r) = \sum_{i=1}^N {1\over h_i^3} \tilde{K}(r,r_i,h_i).
\end{equation}
The optimal way to vary $h_i$ is according to Abramson's (1982) rule:
\begin{equation}
h_i \propto \nu^{-\alpha}(r_i),\ \ \ \ \alpha = 1/2.
\end{equation}
Since we don't know $\nu(r)$ a priori, we instead varied $h_i$ according 
to
\begin{equation}
h_i \propto r^{\beta}.
\end{equation}
We found that $\beta=1/2$ gave good results, which
is reasonable since density profiles are close to
$\nu\sim r^{-1}$.
One could improve on this by first constructing a pilot estimate
of $\nu$ then using Abramson's rule.

The surface density profile could be computed via simple projection of
$\hat\nu(r)$.  Instead, we computed $\hat\Sigma(R)$ directly from the
coordinates projected along one axis.  The two-dimensional kernel
estimate of $\Sigma({\bf R})$ in the absence of any symmetries is
\begin{equation}
\hat\Sigma({\bf R}) = \sum_{i=1}^N {1\over h^2} K'\left[{1\over h}
\left|{\bf R} - {\bf R}_i\right|\right]
\end{equation}
where $K'$ is the two-dimensional Gaussian kernel,
\begin{equation}
K'(y) = {1\over 2\pi}e^{-y^2/2}.
\end{equation}
Now smear each particle uniformly in angle $\phi$ at fixed $R_i$.
The density estimate becomes
\begin{mathletters}
\begin{eqnarray}
\hat\Sigma(R) &=& \sum_{i=1}^N {1\over h^2} {1\over 2\pi} \int
K'\left({d\over h}\right) d\phi, \\
d^2 &=& R_i^2 + R^2 - 2RR_i\cos\phi .
\end{eqnarray}
\end{mathletters}
In terms of the angle-averaged kernel $\tilde{K'}$:

\begin{mathletters}
\begin{eqnarray*}
\hat\Sigma(R) &=& \sum_{i=1}^N {1\over h^2} \tilde{K'}(R,R_i,h),\\
\tilde{K'}(R,R_i,h) &\equiv&  
{1\over 2\pi} \int K' \nonumber\\
&& \left(h^{-1}\sqrt{R_i^2 + R^2 - 2RR_i\cos\phi}\right) d\phi \\ 
& =&{1\over 2\pi} e^{-(R_i^2+R^2)/2h^2}I_0(RR_i/h^2)
\end{eqnarray*}
\end{mathletters}

where the last expression was derived using the Gaussian kernel; $I_0$
is the modified Bessel function.

\begin{figure}
\begin{center}
\epsfig{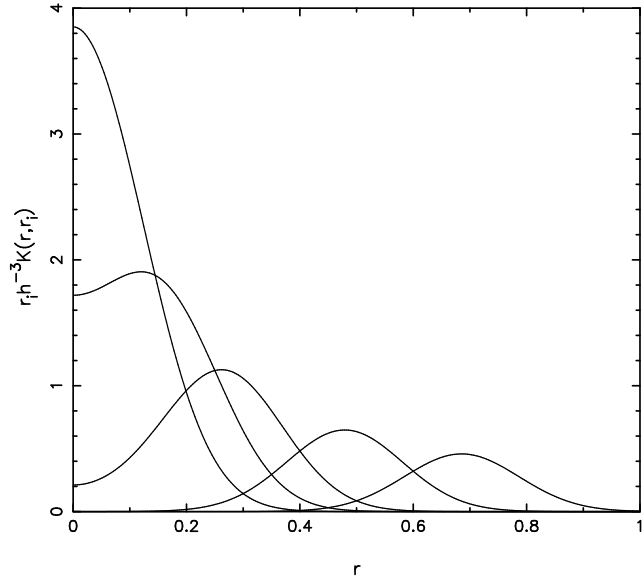}
\caption{Kernels for the radial density estimation problem. Window width is
$h=0.1$ and particles are located at r$=(0.1,0.2,0.3,0.5,0.7)$.}
\label{appfig}
\end{center}
\end{figure}

\label{lastpage}


\begin{thebibliography}{}

\bibitem{abram} Abramson, I. S. 1982, Ann. Stat. 10, 1217

\bibitem{bard} Bardeen, J.M., Bond, J.R., Kaiser, N., Szalay, A.S., 1986,
ApJ, 305, 15.

\bibitem{wmap} Bennett, C. L. \etal, 2003, ApJS, 148, 97B

\bibitem{kdtree} Bentley J. L., 1975, Communication of the ACM 18, 9

\bibitem{bullockpros} Bullock J. S., Kolatt T. S., Sigad Y., Somerville,
R. S., Kravtsov A. V., Klypin A. A., Primack J. R., Dekel A., 2001a,
MNRAS, 321, 559

\bibitem{bullunivmom} Bullock J. S., Dekel, A., Kolatt T. S., Kravtsov A.
V., Klypin A. A., Porciani, C., Primack, J. R., ApJ, 2001b, 555, 240

\bibitem{coleplaw} Cole S., Lacey C., 1996, MNRAS, 281, 716

\bibitem{croneplaw} Crone M., Evrard A., Richstone D., 1994, ApJ, 434, 402

\bibitem{davis} Davis, M., Efstathiou, G., Frenk, C.S., White, S.D.M.,
1985, ApJ, 292, 381

\bibitem{blokrot} de Blok W. J., McGaugh S. S., Bosma A., Rubin V. C.,
2001, ApJ, 552, L23

\bibitem{zurich2body} Diemand J., Moore B., Stadel J., Kazantzidis S., 2004, 
MNRAS, 348, 977

\bibitem{ekeproconc} Eke V. R., Navarro J. F., Steinmetz M., 2001,
ApJ 554, 114

\bibitem{1996MNRAS...282..263}
Eke, V.R., Cole, S., Frenk, C.S., 1996, MNRAS, 282, 263

\bibitem{evans} Evans N. W., Ferrer F., Sarkar S., 2003, PhRvD, 69, 123501

\bibitem{flprimrot} Flores R. A., Primack J. R., 1994, ApJ, 427, L1

\bibitem{fukupro30m} Fukushige T., Kawai A., Makino J., 2004, ApJ, 606, 625

\bibitem{fukusteep}
Fukushige, T., \& Makino, J. 1997, ApJ, 477, L9

\bibitem{fuku1}
Fukushige, T., \& Makino, J. 2001, ApJ, 557, 533

\bibitem{fuku2}
Fukushige, T., \& Makino, J. 2003, ApJ, 588, 674

\bibitem{clustlenincon} Gavazzi R., Fort B., Mellier Y., Pello R., 
Dantel-Fort M., 2003, A\&A, 403, 11

\bibitem{5spirals} Gentile G., Salucci P., Klein U., Vergani D., Kalberla P,
2004, MNRAS, 351, 903

\bibitem{Ghignasub1} Ghigna S., Moore B., Governato F., Lake G., Quinn T.,
Stadel J., 1998, MNRAS, 300, 146

\bibitem{ghignasub2} Ghigna S., Moore B., Governato F., Lake G., Quinn T., Stadel J.,
2000, 544, 616

\bibitem[Governato, Ghigna, \& Moore(2001)]{2001aats.conf..469G}
Governato, F., Ghigna, S., \& Moore, B.\ 2001, ASP Conf.~Ser.~245:
Astrophysical Ages and Times Scales, 469

\bibitem{hayashipro} Hayashi, E., \etal, 2003, astro-ph/0310576

\bibitem{hernpro} Hernquist L., 1990, ApJ, 356, 359

\bibitem{neffpro} Hoffman Y., Shaham J., 1985, ApJ, 297, 16

\bibitem{huffprocneff} Huffenberger K. M., Seljak U., 2003, MNRAS, 340, 1199

\bibitem{huss} Huss A., Jain B., Steinmetz M., 1999, ApJ, 517, 64

\bibitem{jim} Jimenez R., Verde L., Oh P., 2003, MNRAS, 339, 243

\bibitem{jing1}Jing, Y. P., \& Suto, Y. 2000, ApJ, 529, L69

\bibitem{jings2}Jing, Y. P., \& Suto, Y. 2002, ApJ, 574, 538


\bibitem{katzrenormoverm} Katz N., White S., 1993, ApJ, 412, 478

\bibitem{klyppro}
Klypin, A., Kravtson, A. V,, Bullock, J. S., \& Primack, J. R. 2001,
ApJ, 554, 903

\bibitem{lacey94} Lacey C., Cole S., 1994, MNRAS, 271, 676

\bibitem{lakenumass} Lake, G., 1989, AJ, 98, 1253

\bibitem{steepxraypro} Lewis A. D., Buote D. A., Stock J. T., 2003,
ApJ, 586, 135

\bibitem{merrittprokern} Merritt D., Tremblay B., 1994, AJ, 108, 514

\bibitem{merrittwww} Merritt, D. 1994, \newline
http://www.rit.edu/$\sim$drmsps/inverse.html

\bibitem{moorerot} Moore B., 1994, Nature, 370, 629

\bibitem{m98} Moore B., Governato F., Quinn T., Stadel J., Lake G., 1998, AJ, 499, L5 
(M98)

\bibitem{1999MNRAS.310.1147M} Moore B., Quinn T., Governato F., 
Stadel J., \& Lake G., 1999 (M99), MNRAS, 310, 1147

\bibitem{mooredwarf} Moore B., Calcaneo-Roldan C., Stadel J., Lake G., 
Sebastiano G., Governato G., 2001, Phys. Rev. D, 64, 063508

\bibitem{nfw1} Navarro J. F., Frenk C. S., White S. D. M., 1996, ApJ, 462, 563 (NFWa)

\bibitem{nfw2} Navarro J. F., Frenk C. S., White S. D. M., 1997, ApJ, 490, 493 (NFWb)

\bibitem{juliopro} Navarro J. F., \etal, 2004, MNRAS, 349, 1039

\bibitem{powerres} Power, C., Navarro, J. F., Jenkins, A., Frenk, C. S.,
White, S. D.  M., Springel, V ., Stadel, J., \& Quinn, T., 2003, MNRAS,
338, 14

\bibitem{ps} Press W.H., Schechter P., 1974, ApJ,
187, 425

\bibitem{reed2003} Reed, D., Gardner, J., Quinn, T., Stadel, J., Fardal, M.,
Lake, G., \& Governato, F., 2003, MNRAS, 346, 565

\bibitem{ricottipro} Ricotti M., 2003, MNRAS, 344, 1237

\bibitem{salucci} Salucci P., Burkert A., 2000, ApJ, 537, L9

\bibitem{salucci3} Salucci P., 2003, astro-ph/0310376

\bibitem{sand1} Sand D. J., Treu T., Ellis R. S., 2002, ApJ, 574, L129

\bibitem{sand2} Sand D. J., Treu T., Smith G. P., Ellis R. S., 2004, ApJ, 604,
88

\bibitem{scott} Scott, David W. 1992, "Multivariate Density Estimation", 
John Wiley and Sons, New York, p. 130

\bibitem{flatclustlensnot} Shapiro P. R., Iliev I. T., 2000, ApJ, 542L, 1

\bibitem{2drotcurve} Simon J., Bolatto A., Leroy A., Blitz L., 2003, ApJ, 
596, 957

\bibitem{wmapn1} Spergel D., \etal, 2003, ApJS, 148, 175

\bibitem{splinter} Splinter R., Melott A., Shandarin S., Suto Y., 1998, ApJ,
497, 38

\bibitem{stadel} Stadel, J, 2001, PhDT.

\bibitem{stoehr} Stoehr F., White S., Springel V., Tormen G., \& Yoshida N.,
2003, MNRAS, 345, 1313

\bibitem{subrapro} Subramanian K., Cen R., Ostriker, J. P., 2000, ApJ,
538, 528

\bibitem{syerwhiteunivpro} Syer D., White S. D. M., 1998, MNRAS, 293, 337

\bibitem{tasitsiomi} Tasitsiomi A., Kravtsov A., Gottlober S., Klypin A., 2004,
ApJ, 607, 125

\bibitem{taylornav} Taylor J., Navarro J., 2001, ApJ, 563 483

\bibitem{tremaingunnphase} Tremaine S., Gunn J., 1979, Phys. Rev. Lett.
42, 407

\bibitem{tysonsflatlens} Tyson J. A., Kochanski G. P., Dell'Antontio I.
P. 1998, ApJ, 498, L107

\bibitem{frankbeam1} van den Bosch F. C., Robertson B. E. Dalcanton J. J.,
de Blok W. J. G., 2000, AJ, 119, 1579

\bibitem{frankbeam} van den Bosch F., Swaters R., 2001, MNRAS, 325, 1017

\bibitem{gas} Wadsley J., Stadel J., Quinn T., 2004, NewA, 9, 137

\bibitem{risaconc} Wechsler R. H., Bullock J. S., Primack J. R., 
Kravtsov A. V., Dekel A., 2002, ApJ, 568, 52.

\bibitem{zhao} Zhao H. S., 1996, MNRAS, 278, 488

\end{thebibliography}
\end{document}